# An interstellar communication method: system design and observations

*William J. Crilly Jr.*

Green Bank Observatory, West Virginia, USA

*Abstract*— **A system of synchronized radio telescopes is utilized to search for hypothetical wide bandwidth interstellar communication signals. Transmitted signals are hypothesized to have characteristics that enable high channel capacity and minimally low energy per information bit, while containing energy-efficient signal elements that are readily discoverable, distinct from random noise. A hypothesized transmitter signal is described. Signal reception and discovery processes are detailed. Observations using individual and multiple synchronized radio telescopes, during 2017–2021, are described. Conclusions and further work are suggested.**

## I. INTRODUCTION

Interstellar communication signals are hypothesized to provide high information channel capacity and minimal energy expended per information bit transmitted. Assuming the signals are intended by the transmitter to be discovered, at least one discovery mechanism is expected to be included in the transmitted signal design. Information is hypothesized to be communicated on channels that utilize signal polarization, as one means of multiplexing information transfer, and to aid discovery.

The information transfer capacity of a channel is described by the communication principles explained by Shannon [1].

A high information capacity signal transmitted under the constraints of the Shannon limit of energy per information bit, given a level of noise power spectral density, is indistinguishable from Additive White Gaussian Noise (AWGN) of the same average power, each measured within the occupied bandwidth of the signal [1]. The discovery of such optimum signals is therefore problematic, absent a signal discovery mechanism, designed and implemented within the signal [3].

In the present work, a signal discovery mechanism is hypothesized to comprise infrequent, narrow-band elements of the modulated information-bearing signal, transmitted at a higher power level, compared to other elements of the signal, resulting in an aperiodic higher received power level of infrequent narrow-band elements, within an otherwise wide bandwidth communication signal. The possibility of signal discovery is enhanced due to the statistical properties of the narrow-band elements, referred to, in this work, as $\Delta t\ \Delta f$ discovery signals and $\Delta t\ \Delta f$ pulse pairs.

The remainder of this paper is organized as follows. A hypothesis is presented. In the Transmitter Design section, the design of hypothesized interstellar transmission signals is detailed. In the Receiver Design section, the experimental system of radio telescopes, receivers, post-processing and RFI amelioration is described, including metrics to estimate event likelihood in AWGN. The Observations section covers experimental results obtained from 2017–2021 observations, using multiple radio telescopes, and includes calculations of the likelihood of observations in AWGN. Discussion and conclusions follow, and further planned work is described.

## II. HYPOTHESIS

Narrowband signal discovery elements are hypothesized to be contained within frequency and time constrained bundles, in energy bursts, designed for propagation within an interstellar coherence hole (ICH), largely free of propagation impairments, described by Messerschmitt [2], and within coherent signals concentrated in frequency or time, described by Oliver [3].

Assuming that multiplexing methods may be utilized by the transmitter to enhance composite communication channel capacity, the proposed discovery mechanism is hypothesized to function in the presence of channel multiplexing methods, including spatial multiplexing and polarization modulation [3].

The experiment described in this paper tests a hypothesis, stated as follows:

*Hypothesis:* Experimental observations of narrow bandwidth ICH-constrained elements of hypothetical interstellar transmitted signals, are expected to be explained by an AWGN source model, while differing polarization radio telescope receivers, and geographic-spaced synchronized radio telescopes, are pointed to celestial coordinates 5.25 ± 0.15 hours Right Ascension (*RA*) and -7.6° ± 1° Declination (*DEC*).

Falsification of the hypothesis ensues if likelihood functions, derived from an AWGN source model, calculate posterior probability values that are anomalously low, while differing polarization radio telescope receivers, and geographic-spaced synchronized radio telescopes are pointed to celestial coordinates 5.25 ± 0.15 hours *RA* and -7.6° ± 1° *DEC*.

The uncertainty ranges of the *RA* and *DEC* in the hypothesis were derived from investigative pre-hypothesis observations and telescope metrology measurements during 2017 and 2018. The *DEC* center value was set from investigative pre-hypothesis observations, pointing off, and pointing on, 40 Eridani on May 26, 2017, observing with the Green Bank Telescope, through the Breakthrough Listen project [4], observing with the Green Bank Forty Foot Educational Telescope starting in 2015, and beam transit observing at various *DEC*s with radio telescopes in New Hampshire, starting in 2013. The *RA* center value was set after two pairs of $\Delta t = 0$, $\Delta f \approx 0$ Hz pulse anomalies were observed, using the Forty Foot Educational Telescope at Green Bank, synchronized with the Plishner Sixty Foot Telescope near Haswell, Colorado, when pointing the telescopes near 5.25 hours *RA*, -7.6° *DEC* on August 15 and 16, 2018.

The experimental hypothesis is believed to have a high level of falsifiability, an important experimental objective, explained by Popper [5].

Experiments that estimate the likelihood of observation indications relevant to other hypotheses, and their models, e.g. Radio Frequency Interference (RFI), astronomical natural sources, receiving equipment, statistical errors, scintillation, experimental bias, intentionally transmitted interstellar signals, are topics of ongoing and future work, and not addressed in this paper.

## III. TRANSMITTER DESIGN

### A. *Hypothesized discoverable transmitted signals*

If a transmitting entity and receiving entity mutually accept the discovery principles described in **I. INTRODUCTION, II. HYPOTHESIS**, and the rationale explained below, then it seems plausible that the transmitting entity might design a transmitting system to use means to decrease the AWGN-model likelihood of intentionally discoverable signals.

A high information rate and energy-efficient transmitting system utilizes wide bandwidth, and naturally contains aperiodic narrowband signal elements having interarrival time, $\Delta t$, and interarrival RF frequency, $\Delta f$, at significantly lower values than the average interarrival times and average interarrival RF frequencies of elements of the full transmitted signal. These anomalous, amplitude-outlier narrow-band elements are increased in amplitude by the transmitter to indicate, when received in polarization-specific filters having a matched integration time, the presence of an information-bearing, intentionally discoverable signal.

Rationale of the hypothesis that a transmitter uses elements of a wideband communication signal itself, to convey $\Delta t$ $\Delta f$ discovery signals, follows.

**Reduce discovery signal average power penalty**: The penalty in transmitter energy efficiency, due to the transmission of higher average power discovery signals, may be minimal, because the signal elements transmitted at a higher average power level occur infrequently within a communication signal.

**Reduce interstellar transmitter-caused interference**: The $\Delta t$ $\Delta f$ discovery signal mechanism may be used by the transmitter during absences of wide bandwidth communication signals directed towards a potential receiver. These discovery signals have a low duty cycle and are expected to produce a low level of interference to other communication signals. The transmitter is able to transmit discovery signals in many directions, while utilizing most of the transmitter average power transmitting high information content signals in certain directions. The information content in discovery signals may be learned by nearby receivers, ameliorating interference to these receivers.

**Convey polarization and matched filter integration time:** Increased amplitude of discoverable $\Delta t$ $\Delta f$ elements convey to the receiver knowledge of the communication signal's intended matched filter integration time $T$, and quantized polarizations, hypothetically utilized during time intervals that the energy bursts of discoverable signals are transmitted. Receivers do not need to search possible matched filter integration times, and polarizations, because the randomness of a wideband high-capacity signal results in multiple aperiodic $\Delta t$ $\Delta f$ elements received, having nearly equal matched filter integration time.

**Increase Discovery Rate:** Augmented back-end receiver processing allows discovery rate to be enhanced, while $N_{POL}(N_{POL}$ -1) / 2 $\Delta t$ $\Delta f$ discovery signal measurements are facilitated, where $N_{POL}$ is the number of receiver polarization channels.

**Convey signal occupied bandwidth:** The highest and lowest measured RF frequencies of discovered $\Delta t$ $\Delta f$ elements may be used by the receiver to estimate the occupied bandwidth of a decodable wide bandwidth communication signal. Knowledge of occupied bandwidth is helpful in the amelioration of RFI.

**Facilitate a decoding process:** A communication signal decoding process is facilitated, due to the presence of $\Delta t$ $\Delta f$ discovery elements within the communication signal, deduced to have been received at a lower symbol error rate than those of the full communication signal.

**Reduce shared information required**: Increasing the amplitude of existing signal elements, in a measurable way, seems to introduce a minimal amount of information shared by the transmitter and receiver, and needed for communication signal discovery and subsequent decoding.

**Reduce interference to the wideband signal:** The value of the discovery signal's amplitude increase may be measured and accurately estimated, and applied to the received signal, to restore the original transmitted signal, without adding the ambiguity of information content in the $\Delta t$ $\Delta f$ discovery elements interfering with wideband signals.

**Convey estimated transmission impairments**: The density of measured values of $\Delta t$ and $\Delta f$ provides the receiver with information about the propagation impairments that the transmitter estimates to be present in a communication channel that effectively utilizes the ICH. Shared transmitter-receiver knowledge of propagation impairments simplifies signal decoding.

**RFI excision processes are facilitated:** The transmitter's selection of discoverable $\Delta t$ $\Delta f$ elements within a wide bandwidth transmission entails the repetition of pulses having infrequent $\Delta t$ and $\Delta f$ components, at a temporary value of matched filter integration time. Receivers are then able to design RFI excision algorithms that reject signals that do not have these properties. Discoverable $\Delta t$ $\Delta f$ elements have properties that are closer to AWGN properties, than to the properties of known or suspected RFI, in astronomical protected frequency bands. Further discussion of RFI excision is in a section below, **V. RECEIVER DESIGN** *F. Radio Frequency Interference (RFI) Amelioration*.

**Doppler de-spreading is not required for discovery:** Hypothetical discoverable transmitted signals are designed to have matched filter integration times contained within the ICH, and therefore do not require compensation of propagation impairments due to transmitter and/or receiver acceleration along the propagation path.

**Facilitate Angle-of-Arrival (AOA) measurements:** A high information capacity communication signal utilizes wide bandwidths, and lends itself to improved accuracy of AOA measurements, using receive antenna arrays arranged on one or more baselines. Discovery signals that are constrained to a relatively limited RF frequency range, $\Delta\omega$, either due to limited transmitter occupied bandwidth, or receiver instantaneous bandwidth, provide a limited density of measurements along the line defined by $\tau_g = \partial\varphi/\partial\omega$ where $\tau_g$ is the difference in geometric path delay between array elements, $\varphi$ is the measured signal phase difference between array elements, and $\omega$ is the measured frequency of the signal, within the frequency range defined by $\omega = \omega_0 \pm \Delta\omega/2$. Extrapolating measured phase and frequency to obtain an estimate of $\tau_g$ introduces ambiguity due to phase wrapping, while the uncertainty in $\tau_g$ may be proportional to $\Delta\omega^{-1/2}$ or $\Delta\omega^{-3/2}$ depending on how $\Delta\omega$ is distributed in bands [6]. Accurate AOA measurements compel the use of wide overall bandwidths, potentially covering multiple bands.

Accurate AOA measurements of discovery signals reduce the time required to receive high information capacity transmitters, expected to have lower received flux on antenna apertures, measured in $W \cdot m^{-2} \cdot Hz^{-1}$, compared to discovery signals. Discovery signals are therefore expected to utilize overall high occupied bandwidth.

**Simplify discovery receiver design:** Discovery receivers may be placed at various locations on a planet, to ameliorate RFI and increase discovery system duty-cycle. Data from these receivers is then combined for post-processing. Data archiving and data transmission issues compel a mechanism to reduce the data required for discovery. The transmission of high average power, short duration discovery signals might simplify a distributed post-processing discovery mechanism.

**Receiver cooperation may be important to transmitting civilizations:** Transmitting civilizations may have concern about technology-nascent civilizations self-harming due to the reception of interstellar signals. The use of a discovery signal mechanism that implies a need for many receivers, covering many areas of receiving planets, together with a relatively slow discovery and decoding process, may be a message that receiver cooperation is encouraged.

**Convey the use of spatial multiplexing:** Transmitting civilizations may have implemented spatial channel multiplexing to increase the multichannel information bit rate. The use of a discovery mechanism that implies a need for widely spaced receiver antennas, may be a suggestion that spatial multiplexing has been utilized.

### B. Description of a hypothesized transmitter signal

The hypothesized transmitted signal is expressed as follows, in the form of example right-hand and left-hand circularly polarized multiplexed channels, having signal amplitude values $R(t)$ and $L(t)$:

$$R(t) = R_0(t) + A_R(t) \, R_+(t) \qquad (1)$$

$$L(t) = L_0(t) + A_L(t) \, L_+(t) \qquad (2)$$

where:

$R(t)$ is the composite right-hand polarized transmitted signal amplitude,

$R_0(t)$ are the original right-hand polarized signal components to be transmitted at the nominal amplitude,

$A_R(t)$ is a time-varying dimensionless factor of right-hand amplitude increase, $A_R(t) \geq 1$, and

$R_+(t)$ are the original right-hand polarized signal components, having minimal likelihood in AWGN, that are chosen to be increased in amplitude in the right-hand polarized transmitter.

$L(t)$, $L_0(t)$, $A_L(t)$, and $L_+(t)$ correspond to the left-hand polarization-multiplexed signals, and their components, as described above for the right-hand polarized signal.

A designed interval of the transmitted waveform, $T$, is optimally chosen to provide a transmitted signal that has an unchanging combination of basis functions, during the time duration $T$ that a receiver polarization channel's matched filter cross-correlates the received signal with each basis function, and indicates a likelihood of the presence of a basis function, distinct from AWGN.

The transmitted signal hypothesis in this work includes discovery signals that are transmitted at various values of $T$, to avoid the requirement that the receiver search through numerous $T$ values, each corresponding to a matched filter integration time.

An example of $R(t)$ and $L(t)$, while transmitting a discovery mechanism, i.e. $A_R(t) > 1$, and/or $A_L(t) > 1$, comprises the sinusoidal orthonormal basis functions [7], with $R_+(t)$ and $L_+(t)$ each comprising one of two basis functions that have reduced RF frequency difference, measured during the time interval between $t_A$ and $t_A + \Delta t + T$, where $t_A$ is the time at which an anomalous event initiates, $\Delta t$ is the time interval between the initializations of two anomalous events, and $T$ is the intended matched filter integration duration. Equations (3) – (6) describe an example.

$R_0(t)$ and $L_0(t)$ comprise the remaining utilized basis functions, absent the signal elements comprising the two anomalous amplitude basis functions.

Basis functions other than orthonormal sinusoidal functions might be chosen by the transmitter, given that their use will produce aperiodic $\Delta t\ \Delta f$ elements that may be similarly increased in amplitude. However, orthonormal sinusoidal basis functions provide the natural framework within the mathematical description of RF frequency quantization, and the $\Delta f$ component of $\Delta t\ \Delta f$ discovery elements. In addition, matched filtering of orthonormal sinusoidal basis functions is readily computed using low complexity algorithms.

Signal element amplitude modulation is expressed in four cases, as examples, as follows, depending on which polarization channel, right-hand, or left-hand, initiates and terminates the amplitude modulation.

Right hand followed by left hand:
$$A_R(t) > 1 \text{ if } \quad t_A \ \le \ t \ \le t_A + T$$
$$A_L(t) > 1 \text{ if } \quad t_A + \Delta t \ \le \ t \ \le t_A + \Delta t + T$$
$$A_R(t) = A_L(t) = 1 \text{ at other times,} \qquad (3)$$

Left hand followed by right hand:
$$A_L(t) > 1 \text{ if } \quad t_A \ \le \ t \ \le t_A + T$$
$$A_R(t) > 1 \text{ if } \quad t_A + \Delta t \ \le \ t \ \le t_A + \Delta t + T$$
$$A_L(t) = A_R(t) = 1 \text{ at other times,} \qquad (4)$$

Right hand followed by right hand:
$$A_R(t) > 1 \text{ if } \quad t_A \ \le \ t \ \le t_A + T$$
$$A_R(t) > 1 \text{ if } \quad t_A + \Delta t \ \le \ t \ \le t_A + \Delta t + T$$
$$A_R(t) = 1 \quad \text{ at other times, and}$$
$$A_L(t) = 1, \qquad (5)$$

Left hand followed by left hand:
$$A_L(t) > 1 \text{ if } \quad t_A \ \le \ t \ \le t_A + T$$
$$A_L(t) > 1 \text{ if } \quad t_A + \Delta t \ \le \ t \ \le t_A + \Delta t + T$$
$$A_L(t) = 1 \quad \text{ at other times, and}$$
$$A_R(t) = 1, \qquad (6)$$

where $t_A$ is the time a $\Delta t\ \Delta f$ burst initiates, having increased amplitude, $\Delta t$ is the time elapsed between a first and second element of the burst, and $T$ is the intended receive matched filter integration duration. **Figure 1** illustrates the spectral and temporal content of hypothetical discovery signals having orthogonal circular polarizations.

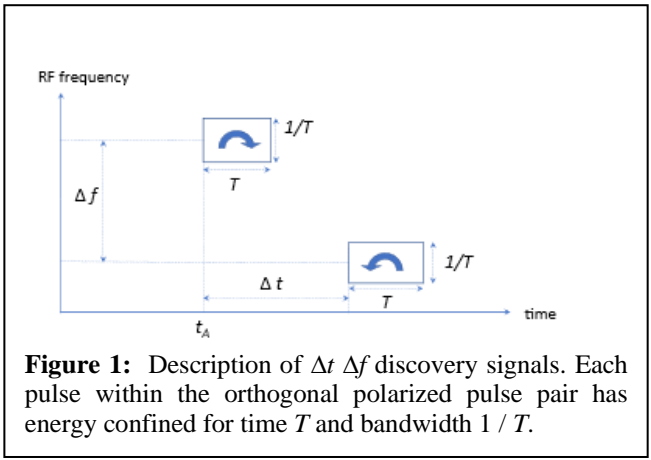

**Figure 1:** Description of $\Delta t\ \Delta f$ discovery signals. Each pulse within the orthogonal polarized pulse pair has energy confined for time $T$ and bandwidth $1 / T$.

The subscripts $L$ and $R$ in the $A_{(POLARIZATION)}$ values in (1) through (6) correspond to two values, among a hypothesized set of quantized polarization values, distributed on the Poincaré sphere with Stokes vector space values [8].

The choice of polarization multiplexing in the development of the $A_{(\cdot)}(t)$ and $A_{(\cdot)}(t)$ factors, is one choice among other types of multi-channel multiplexing methods, e.g. spatial multiplexing, using geographic receiver separation, and large transmitter antenna element spacing, and/or intentional transmitter signal scattering and diffraction.

The value of the factors $A_{(\cdot)}(t)$ are expected to be chosen at the transmitter based on the maximum interstellar distance at which the transmitted composite signal is designed to be discovered, together with an estimate of the discovery signal receiver's antenna effective area, relative to an estimate of the composite signal receiver's antenna effective area, while all else is appropriately optimized for discovery and signal decoding.

Discoverable signals are expected to be subsequently decoded, leading to the requirement of a decoding mechanism. At least some of the decoding mechanism is speculated to be transmitted within the discovery mechanism, at time intervals that may be large, due to their infrequent necessity. Further consideration of signal decoding mechanisms is outside the intended scope of this paper.

## IV.    RECEIVER DESIGN

### A.    *Receiver Design Objective*

The receivers used in this experiment are designed to measure the AWGN-model likelihood of the reception of hypothesized signals, as described in the experimental hypothesis, and **III. TRANSMITTER DESIGN**, while excising as much suspected RFI as possible.

Excising RFI is a risky process as it inherently reduces the receiver's ability to discover any signals. Fortunately, the signals hypothesized in this work are similar to AWGN, except in their short term aperiodic $\Delta t\ \Delta f$ characteristics, compelling the use of robust RFI excision algorithms. This apparent advantage is not without the risk of false alarms. Follow-up mechanisms are needed to measure the likelihood that candidate received signals are correlated with excised RFI. In the current work, this follow-up mechanism involves examining RFI excision output files of machine post-processing, and examining files containing machine RFI excision parameters correlated to measured anomalous $\Delta t\ \Delta f$ events. Metrics and algorithms to quantify this likelihood are a topic of ongoing and future work.

In general, a receiver designed to receive a particular type of signal will eventually indicate, by chance alone, the reception of the hypothesized signal, when only AWGN is applied to the receiver. Determining the possible presence of hypothetical signals, in an attempt to refute an AWGN hypothesis, compels the calculation of the AWGN-model likelihood of observations, while assuming that an AWGN-only model explains the anomalous events.

AWGN-model likelihood, used in the context of this paper, applies only within the constraints of the particular model, assumptions, hyperparameters, and data populations relevant to the particular calculation. Calculated likelihoods are not intended to imply significance of data in other contexts.

AWGN-model likelihood functions, relevant to the measurement of polarization, signal-to-noise ratio (SNR), $\Delta t$ $\Delta f$ energy bursts and *RA* are described in the **APPENDIXES.**

### B.    Radio telescopes

In this experimental work, three radio telescopes are utilized: the Forty Foot Educational Telescope of the Green Bank Observatory, in Green Bank, West Virginia**,** the Plishner Sixty Foot Telescope of the Deep Space Exploration Society, near Haswell, Colorado; and the Twenty-six Foot Telescope in Dunbarton, New Hampshire. The three radio telescopes are in places of low and controlled RFI.

The telescopes' local oscillators and time measurements, are synchronized using stable oscillators locked to GPS satellite signals. Time stamps are processed synchronously with telescope receiver signals, and posted to telescope raw data files. Doppler offsets and metrology-derived frequency corrections are applied during the post-processing of previously stored raw data files.

Polarization reception is linear on the Forty Foot telescope, circular on the Sixty Foot telescope, and dual circular on the Twenty-Six Foot telescope.

Telescopes are pointed to provide overlapping antenna pattern response on the celestial sphere, while beam transit scanning at -7.6° *DEC*. The Forty Foot Telescope at Green Bank has azimuth pointing at 180 degrees. Antenna beam overlap and telescope sensitivity is verified using celestial object NRAO 5690 as a radio calibration source.

### C.    Receivers

Receivers operate within a range of approximately 1395 – 1455 MHz, chosen due to engineering and astronomical considerations [9][3]. The internationally protected band at 1400 – 1427 MHz has remained relatively free of intentionally transmitted signals and unintentionally transmitted energy [10]. On the other hand, the 1427 to 1455 MHz range has numerous potential RFI sources, including an aeronautical telemetry band [11].

Prior to April 2019, receivers used cavity-filtered single sideband down-conversion to an intermediate frequency, digitized at 125 MSPS at 8 bits per sample. The local oscillator was set to 1400 MHz, phase-locked to an oven-controlled crystal oscillator frequency-locked to GPS signals.  In the April 2019 and subsequent observation runs, receivers use in-phase quadrature (IQ) down-conversion, with a GPS-locked local oscillator set to 1425 MHz. The digitizers sample the I and Q channels at 62.5 MSPS at 8 bits per sample. A digitizer data capture is triggered in hardware by a signal from the GPS receiver, set to match a UTC-second time.

Baseband IQ time domain data is transferred to computers over USB 3.0, at approximately 5 Gbps. Fast Fourier Transforms (FFT) are performed using FFTW3 [12] in multi-core processor systems, using POSIX pthreads to synchronize pipelines of sampling, data transfer, calculation, and measurement data storage to a file. New telescope output files are created automatically at four-hour intervals.

Spectral channels in the FFTs of the receivers have a 3.7 Hz bandwidth with an integration time $T = 1 / 3.7$ Hz = 0.27 seconds. Each resulting bandpass filter and post-integration sampler may be considered a filter matched to a narrowband element of transmitted signals.

SNR is calculated for each 3.7 Hz channel during each 0.27 second time interval, and for a composite set of three consecutive 0.27 second time intervals, the latter used to ameliorate sensitivity loss due to pulses straddling consecutive time intervals.

Signal quantities in SNR are measured as the average power in a 3.7 Hz bandwidth, during one, and three consecutive periods of 0.27 seconds, the latter described in the previous paragraph. Signal average power measurement values include the contributions of average noise power. Estimated average noise power is not subtracted from the average signal power in the SNR measurement. SNR is defined as $(S+N)_{MEASURED}/N_{MEASURED}$.

Noise quantities in SNR are measured by averaging the power in a 954 Hz bandwidth, overlapping with the FFT bin frequency, over a time period of four consecutive FFT periods, and reducing the measured power value by the bandwidth ratio factor, to estimate the average noise power in a 3.7 Hz bandwidth at the FFT bin frequency.

Raw data stored to files contains details of signal measurements pertaining to SNR events that exceed 11.8 dB composite three sample $SNR_{COMP}$, and exceed 11.0 dB single sample time interval.

$SNR_{MAX}$ is the $SNR_{COMP}$ value of the maximum of the $SNR_{COMP}$ values of the pair of SNR events that have $\Delta t$ and $\Delta f$ measured. SNR used in this paper refers to $SNR_{MAX}$.

Receivers have implemented an increasing duty-cycle. Haswell and Green Bank receivers increased duty-cycle from 25% to 33% before the April, 2019 observation run. Dunbarton receivers' duty-cycles were increased to 100% in March, 2020. The duty-cycle in this context is short-term, e.g.  25% duty-cycle corresponds to one second reception, three seconds processing. 100% duty-cycle is implemented using three dual channel digitizers, per polarization, each triggered by a signal from a GPS receiver.

Time of day encoded signals are transmitted on audio frequencies from the GPS receivers to one channel of each dual channel digitizer, so that time may be synchronized in hardware with analog receiver baseband signals.

Receiver systems do not communicate with each other, to reduce the risk of corruption of otherwise independent geographically-spaced and polarization-differing signals.

### D.    Observation post-processing

The values of $\Delta t$ and $\Delta f$ measurements are defined in this experiment to be interarrival time, and interarrival frequency of high SNR elements of signals, after Earth-rotation Doppler offset frequency is compensated, if required. Interarrival time differences of ICH-constrained signals received at multiple telescopes are not compensated, as the arrival time differences at the telescopes utilized in this experiment are assumed to be negligible compared to the experiment's 0.27 second matched filter integration time.

 Low valued measurements of the hypothesized signal element pair interarrival time, $\Delta t$, and interarrival frequency, $\Delta f$, of spectral elements exceeding the receiver SNR threshold are expected to be infrequent, in AWGN, and used to establish the AWGN-model likelihood of the $\Delta t$ $\Delta f$ event, quantified in **APPENDIX C** *AWGN-model likelihood of  $\Delta t$ $\Delta f$ elements in hypothesized energy bursts*.

The measurement of high SNR $\Delta t \, \Delta f$ discovery signals, performed in the post-processing of telescope receiver raw data files, compares a pair of telescope raw-data files, comprising measurements made at the time of each SNR threshold crossing event. Each file within the file pair corresponds to a geographical location, and/or a differing polarization sense, and spans a four-hour time interval.

Machine post-processed $\Delta t \, \Delta f$ anomalies are reported in this paper for the geographical telescope pair of Haswell and Green Bank, and the polarization pair comprising co-located Dunbarton orthogonal circular polarizations. Other $\Delta t \, \Delta f$ file pair comparisons, including six pairs of observations comprising three telescopes, one having dual polarization, have been reported in previous presentations [13] [14].

$\Delta t \, \Delta f$ anomalies have been observed to coincide in time with $\Delta t = 0$, $\Delta f \approx 0$ Hz anomalies. The former anomalies are referred to as associated pulse pair anomalies, and defined by their presentation at the same Modified Julian Date (MJD) as a $\Delta t = 0$, $\Delta f \approx 0$ Hz anomaly.

Machine post-processing stores the receivers' raw data records within a settable time window, around the MJD of a $\Delta t = 0$, $\Delta f \approx 0$ Hz anomaly. In this way, $\Delta t \, \Delta f$ events that are associated with $\Delta t = 0$, $\Delta f \approx 0$ Hz may be identified and examined for AWGN-model likelihood.

During machine post-processing, the *RA* of the receiving telescope's pointing direction is calculated, based on the MJD of each SNR $\Delta t \, \Delta f$ event, and a 180° azimuth Green Bank reference. The presence of high SNR $\Delta t \, \Delta f$ discovery elements observed at one or more *RA*s, within one or more beam transits, may then be examined.

Receiver raw data files of observation runs are retained so that post-processing having different hyperparameters may be subsequently applied to archived files.

### E. Radio Frequency Interference (RFI) Amelioration

RFI is ameliorated, in current and past observations, using up to nine RFI analysis, amelioration and excision processes. The machine processes $2) - 5)$ described below, are utilized during current (April 2021) machine post-processing, and were designed and implemented as the automation of post-processing increased. Machine process 1) has been utilized throughout all observations.

Observations publicly reported, prior to this paper, [13][14] did not make full use of the machine processes $2) - 5)$. Rather, manual RFI excision was performed during post-processing.

In this paper, all reported observations, from 2018 until the present, use machine processes $2) - 5)$ during post-processing. RFI model hyperparameters, used in the post-processing software, are retained at the same set values, throughout the reported observations, except as indicated in the figures and the text. For example, during three days of the 164 beam transit observations, suspected RFI was manually excised after machine post-processing, and detailed in the presentation of results, in **V. OBSERVATIONS, F. 164 beam transits of the Twenty-six Foot telescope**.

The use of machine-only RFI excision processes is an important objective of this experimental work, as it facilitates experimental repeatability, traceability, AWGN-model likelihood calculations and corroboration.

RFI excision, in general, increases the risk of excising interstellar signals, confounding potential falsification of the AWGN hypothesis. In this experimental work, the false-RFI excision risk is thought to be ameliorated, due to the similarity of AWGN to the hypothetical transmitted signals, together with the intentional excision of many types of human-made RFI, of unknown source. In one interpretation of this concept, robust and generalized RFI excision, without investigations of the signals, might benefit the search for hypothesized discoverable signals. On the other hand, it is inherently difficult to distinguish an energy-efficient interstellar signal from certain human-made signals, without troubleshooting to the RFI itself.

Descriptions of RFI amelioration processes, utilized in this experiment, follow.

1) **Telescope site-specific persistent RFI:** When practical, site specific RFI sources are identified to the transmitting source, using directional antennas and portable spectrum analysis. This is often a tedious process, resulting in the need for excision of persistent apparent-RFI signals of unknown transmitting source.

During receiver installation at each telescope site, time- and frequency persistent narrowband signals are discovered in raw data, at differing telescope celestial pointing directions. Received signals in the raw data measurements are binned to associated 954 Hz wide bands, each a spectral segment of 256 contiguous FFT bins. Persistent signals in these segments are excised from the available receiver spectrum of the receiver at the telescope site. The process of identifying the spectral segments to excise is manually performed at each telescope site at the time of receiver installation. Raw data files and telescope receiver source code may be examined, during post-processing work, to determine the frequency range of these excised spectral segments. This work entails the experimental test of an RFI hypothesis, not a subject within the current reported work.

2) **Machine post-processing persistent RFI:** Within the overall received bandwidth, 954 Hz segments that contain an anomalous count of SNR threshold crossings, compared to the count of estimated noise-caused SNR threshold crossings, during a four-hour time interval, are excised. SNR threshold crossing events that are present within these excised bands are not processed, within the four-hour time interval of the receiver raw data file. Persistent RFI, for each four-hour duration file, is determined and machine excised prior to examining pairs of files for $\Delta t \, \Delta f$ SNR anomalies. A persistent RFI measurement file is saved and may be examined in the testing of a future RFI hypothesis.

3) **Machine post-processing dynamic RFI:** Dynamic, narrow bandwidth RFI, appearing in a telescope's receive signal, is excised using approximately sixty-thousand 954 Hz bandwidth Infinite Impulse Response (IIR) filters, per receiver channel, with each IIR filter updating a spectral filtered average SNR measurement, at the time of an SNR threshold crossing event. The IIR filter mechanism is designed to reject pulses that persist in time and frequency longer than expected within a wide bandwidth energy-efficient communication signal, and longer than expected in AWGN. During the time that the output of an IIR filter exceeds an $SNR_{IIR}$ threshold, the 954 Hz bandwidth of the particular IIR filter, is excised. The IIR filter coefficients are set manually, and held at a fixed value, to reject few true negatives, e.g. sporadic high SNR events, within the exponentially distributed power of AWGN, while not significantly rejecting infrequent, low-valued $\Delta t$, $\Delta f$ non-persistent spectral elements. The IIR filter output values are recorded to a post-processing output file and may be used in the experimental test of a future RFI hypothesis. RFI entering sidelobes of high gain antennas typically presents high variance of measured SNR. The IIR filter mechanism may therefore ameliorate antenna sidelobe RFI.

4) **Machine post-processing harmonic frequency RFI:** Unintentional emissions often appear at harmonics of standard frequencies, e.g. 5 MHz yielding 1405, 1410 MHz, etc. During machine post-processing, signals measured within 25 kHz of a harmonic of 500 kHz are not posted to the post-processing output file, and not used in the measurement of $\Delta t$ and $\Delta f$.

5) **Post-observation machine RFI excision:** Spectrum below 1400.8 MHz, above 1447.0 MHz, and between 1424.0 and 1426.0 MHz were excised from the Twenty-six Foot telescope receiver data, in machine post-processing, due to suspected sporadic RFI observed on several days during the 164 beam transit test's machine post-processing of data from the Twenty-six Foot Telescope.

6) **Geographically spaced synchronized telescopes:** Large telescope spacing of the Green Bank, Haswell and Dunbarton telescopes provides a natural RFI reduction mechanism, to an RFI transmitter distance as high as $10^6$ kilometers, assuming the RFI source is located within the minimum of the Full Width at Half Maximum (FWHM) beamwidths of the telescope antennas. Potential sidelobe response remains. The common line-of-sight intersection altitude between Haswell and Green Bank is approximately 80 km. Terrestrial RFI from a common source is expected to be reduced, due to minimal propagation from a terrestrial or airborne source to the geographically spaced telescopes. Assuming the presence of near-Earth space-based RFI of a single transmitting source, entering telescope side-lobes, $\Delta t$ measured between telescope sites is expected to indicate values less than 0.27 seconds, while $\Delta f$, after Earth-rotation Doppler difference correction, will depend on the velocity and location of the RFI source.

7) **Noise source tests:** Examination of post-processed output files, when AWGN test source(s) are applied to polarized receivers, helps find equipment-caused potential RFI, otherwise undetected.

8) **Polarized signal RFI models:** The discovery signals described in this paper, i.e. low values of $\Delta t$ and $\Delta f$, of a single polarized receiver signal, or a joint pair of polarized signals, may be modeled against known types of RFI, and readily provide a means to calculate the likelihood of an RFI cause, given a particular RFI model and its estimated prior likelihood. Bayesian Inference may be used to estimate the likelihood that data is explained by various RFI models. For example, a terrestrial or space-based, communication system of human design, that is intended to be clandestine, would not be expected to be transmitting narrow bandwidth bundles of energy that are relatively easy to detect. Further, such transmissions seem unlikely to be discovered within an internationally protected frequency band. RFI modeling and hypothesis development is further work.

9) **RA filtering:** Observations that span many days may be used to filter RFI that has periodicity, providing an *RA* spreading mechanism. *RA* filtering was used in the current experiment during the 164 beam transit test, described in **V. OBSERVATIONS, F. 164 beam transits of the Twenty-six Foot telescope.**

### F.  Implementation of $\Delta t$  $\Delta f$, SNR, RA, AWGN-model likelihood calculations

An objective of this experiment is to suggest and test a repeatable set of measurements that optimally indicate the presence or absence of hypothesized $\Delta t$ $\Delta f$ discovery signals. Among many possibilities, measurements are chosen that most readily falsify an AWGN hypothesis.

The set of measurements chosen are $\Delta t$, $\Delta f$, *SNR*, MJD and *RA*,  and are processed as follows.

1. Differing polarization signals, or geographically spaced signals, that exceed an SNR threshold, using a single hypothetical matched filter, are measured. The set of measurements $\Delta t$, $\Delta f$, *SNR*, MJD and *RA* are made using data from geographically spaced receivers, or data at co-located receivers. Anomalously low $\Delta t$, low $\Delta f$ measurements are recorded.

2. At the same MJD of anomalous $\Delta t$ $\Delta f$ signals recorded in step 1, the spectra of the two polarized or geographically spaced receivers are searched for associated low $\Delta t$ and low $\Delta f$ signal elements, in a **Method A**.

3. The least likely anomalous $\Delta t$ $\Delta f$ values from step 1 are selected, and binned to *RA* regions and SNR relevant to the hypothesis, in a **Method B**.

4. Given an AWGN model, a **Method A** and/or **Method B** likelihood function value is calculated. The methods are described in **APPENDIX C: AWGN-model likelihood of $\Delta t$ $\Delta f$ elements in hypothesized energy bursts.** In this experiment, **Method A** is used for geographically spaced receiver measurements, reporting binomial cumulative probability values of $\Delta t = 0$, $\Delta f$ pulse pairs associated with a $\Delta t = 0$, $\Delta f \approx 0$ Hz pulse pair. **Method B** is used for the 164 beam transit test, reporting binomial probability density values of $\Delta t = 0$, $|\Delta f| \leq 400$ Hz signals, as a function of decreasing pulse pair SNR. The pair event probability value estimate is reported within binned *RA* regions.

5. Posterior probabilities are calculated using Bayesian data analysis, potentially supporting, or falsifying, the AWGN hypothesis.

### G.  Machine post-processing hyperparameters

In this experiment, machine post-processing hyper-parameters are set to filter events to a range of $| \Delta t | \leq 3$ seconds, and $| \Delta f | \leq 400$ Hz. In the presentation of results,

and in the calculation of posterior probabilities, only $\Delta t = 0$ pulse pair events are presented.

$\Delta t$ $\Delta f$ associated anomalies reported in previous presentations [13][14] include $\Delta f$ anomalies observed outside the ±400 Hz range and at $|\Delta t| \geq 0$. The ±400 Hz range and $\Delta t = 0$ value were chosen in this work to enhance the observation of signal elements least likely to be observed in AWGN.

Anomalies reported in this work are limited to those pulse pairs expected to have minimum likelihood as a function of $\Delta t$, in AWGN, i.e. $\Delta t = 0$, together with pulse pairs that are associated in time with these anomalies. The associated pulse pairs have $\Delta t = 0$. Associated pulse pairs are explained as follows.

Associated pulse pairs are defined as anomalous ICH energy bursts having an MJD matching the MJD of an underlying $\Delta t = 0$, $|\Delta f| \approx 0$ Hz pulse pair. When identifying associated pulse pairs, a maximum $|\Delta f|$ is set by (14), in **APPENDIX C**, by limiting associated pulse pairs to $p_{0\Delta f\,AWGN} < 0.03$, reducing false positives. The associated pulse pair selection process is described as follows.

The collection of $\Delta t = 0$, $\Delta f$ associated pulse pairs entails an examination of the post-processing machine's entries in a post-processing auxiliary output file, and tests the associated pulse pairs against the post-processing RFI filter hyperparameters, described in **IV. RECEIVER DESIGN, E. Radio Frequency Interference (RFI) Amelioration**, processes 2, 3, and 4.

In the reported results, associated $\Delta t$ $\Delta f$ energy bursts are considered associated if a) their pair $\Delta f$ likelihood in AWGN is less than 0.03, b) they present a $\Delta t = 0$ measurement, c) they have the same MJD as a $\Delta t = 0$, $|\Delta f| \approx 0$ Hz pulse pair, and d) they pass tests using the RFI filter processes referenced in the previous paragraph.

SNR hyperparameters are used to develop the number of trials used in the binomial cumulative probability likelihood function calculation. The number of trials $N_{TRIALS}$ in the associated pulse pair likelihood function is the number of $\Delta t = 0$, $\Delta f$ events having an SNR greater than the lowest SNR among the identified associated pulse pairs. The counting of $N_{TRIALS}$ is graphically shown in **Figures 7, 12 and 13**.
.

## V. OBSERVATIONS

### A. *Observations during 2017–2021*

Two-telescope geographically spaced synchronized $\Delta t$ $\Delta f$ observations were conducted at Green Bank and Haswell, during six observing runs, for a total of 180 hours, starting in November 2017 through April of 2019. Semi-continuous observations were made during the earlier of these geographically spaced observations, increasing to 24 hour per day observations in Feb. 2019.

Two-telescope and three-telescope geographically spaced synchronized $\Delta t$ $\Delta f$ observations were conducted at Green Bank, Haswell, and Dunbarton, during a December 2019 observing run of 110 hours duration.

In August, 2020, the Haswell and Dunbarton telescopes were run synchronously for 76 hours duration.

During 2020 and 2021, the Twenty-six Foot telescope operated with two orthogonal circular polarizations,

continuously producing raw data files, except during maintenance and test, and power outages longer than a few minutes.

### B. *Observations present four $\Delta t = 0$, $\Delta f \approx 0$ Hz anomalies at 5.1–5.4 hours RA, -7.6° DEC telescope pointing, during two geographically-spaced observation runs*

The August 15, 16, 2018 Haswell and Green Bank synchronized observation run presented two anomalous $\Delta t = 0$, $|\Delta f| \leq 0.7$ Hz observations, on two adjacent days, while pointing near 5.18 hours *RA* and 5.25 hours *RA*. Examination of raw data file records, within 20 seconds of these simultaneous pulses, yielded an anomalous number of apparently associated low $\Delta t$ and low $\Delta f$ pulses, reported previously [13][14].

The April 2, 3, 2019 observation run presented a $\Delta t = 0$, $|\Delta f| = 5.8$ Hz measurement, and a $\Delta t = 0$, $|\Delta f| = 15.5$ Hz measurement, together with apparent associated $\Delta t = 0$, $\Delta f$ anomalies. Some of these and other anomalies, were reported previously [13][14].

The past analysis of these observations involved post-processing having manual steps, and therefore may have been affected by experimental bias and/or errors difficult to trace, compelling an increased use of machine-only post-processing in this work.

In the present analysis, the machine post-processing of three observation runs, among the larger number of 2017–2021 observation runs, restricts the consideration of general conclusions regarding the full set of 2017–2021 observations.

The geographically spaced observations are described in the next three sections. Six-month duration beam transit observations are then described.

### C. *August 15, 16, 2018 observations indicate two $\Delta t = 0$, $|\Delta f| \leq 0.7$ Hz anomalies within 5.1–5.4 hours RA, -7.6° DEC telescope pointing*

**Figures 2–5** plot measurements relevant to SNR threshold crossing events during the August 15, 16, 2018 observation run. Each data point within a plot indicates a $\Delta t = 0$, $|\Delta f| \leq 400$ Hz measurement, above an $SNR_{MAX}$ of 12.0 dB, with $SNR_{MAX}$ the higher of the two geographically spaced measured SNRs. Machine post-processing and machine RFI excision of telescope files are used. Plots in **Figures 2–5** present the $\Delta f$, RF Frequency, SNR and MJD vs. *RA* of the $\Delta t = 0$, $|\Delta f| \leq 400$ Hz August 15, 16, 2018 measurements. In **Figure 2** two $\Delta t = 0$, $|\Delta f| \leq 0.7$ Hz Green Bank and Haswell pulse pairs are apparent in the 5.1–5.4 hour *RA* region.

The *RA* range in plots is reduced from 24 hours to 0–7.5 hours.

**Table 1** contains measurements of $\Delta t = 0$, $|\Delta f| \leq 3.1$ Hz highest SNR pulse pairs recorded during the August 15,16, 2018 observation run. Two of the pulse pairs, among the five highest SNRs, presented near equal *RA* values. The SNR of the two pulse pairs measured the highest, and third highest, of the five highest SNRs. Their $|\Delta f|$ values measured the two lowest values. The measurements of these two pulse pairs set the specific *RA* center value in the hypothesis.

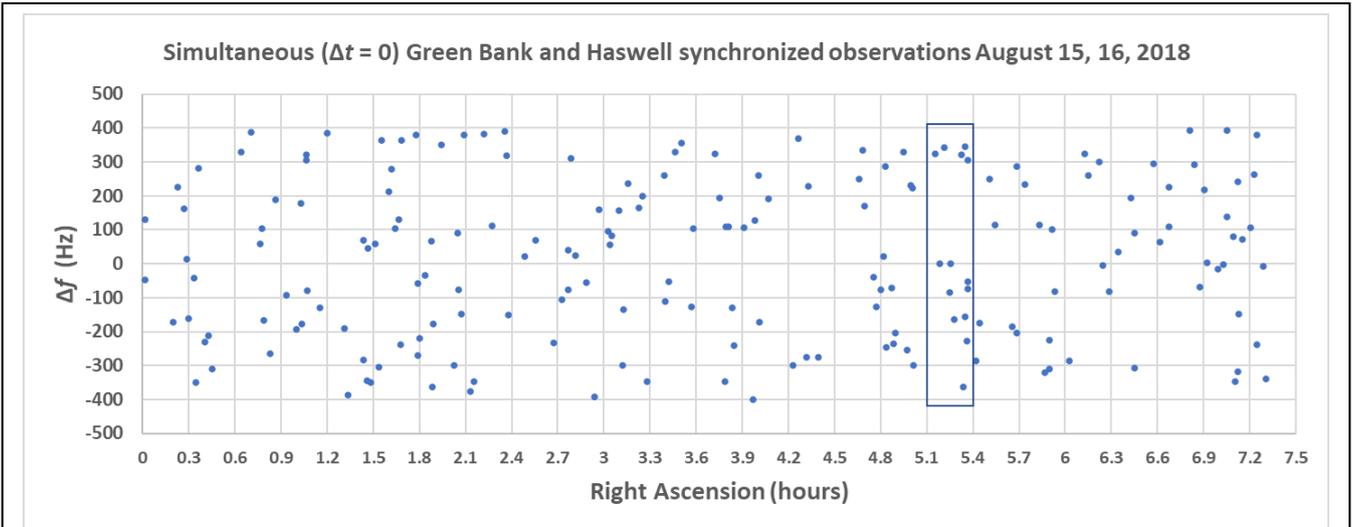

**Figure 2:** Δ*f* **vs** *RA* of pulse pairs during August 15, 16, 2018. Measurements of two near-zero Δ*f* values, at Δ*t* = 0, in a 5.1–5.4 hour *RA* range, are observed.  SNR > 12.0 dB.

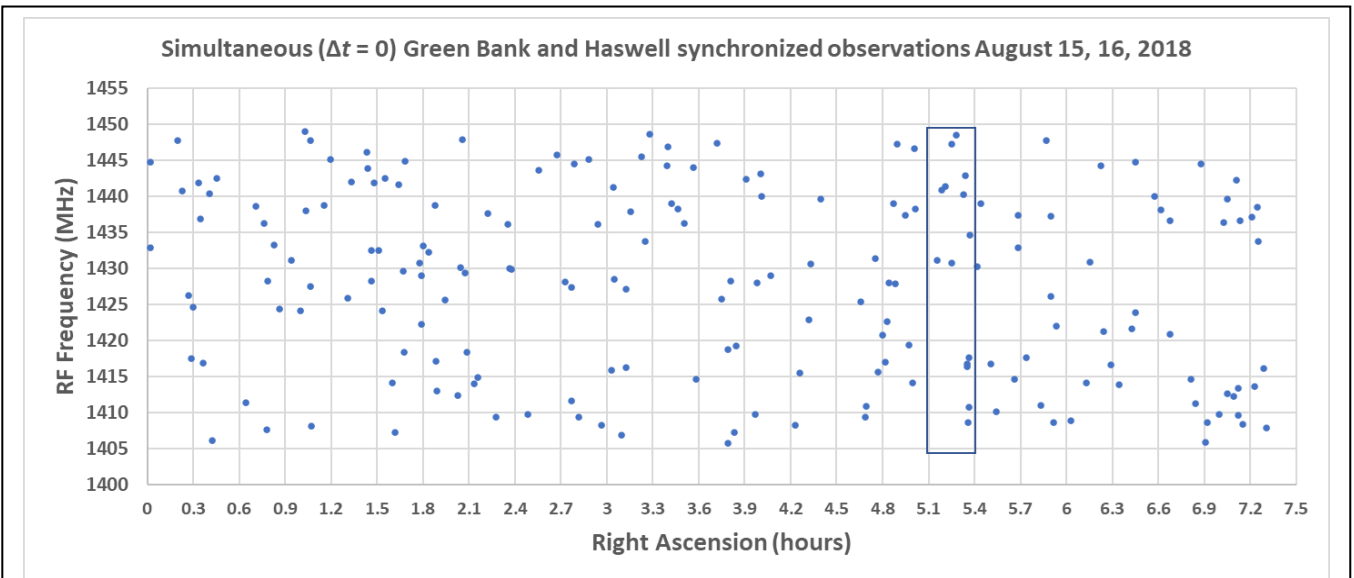

**Figure 3:  RF Frequency vs.** *RA* of Green Bank and Haswell pulse pairs during August 15, 16, 2018.  SNR > 12.0 dB, Δ*t* = 0, | Δ*f* | ≤ 400 Hz.

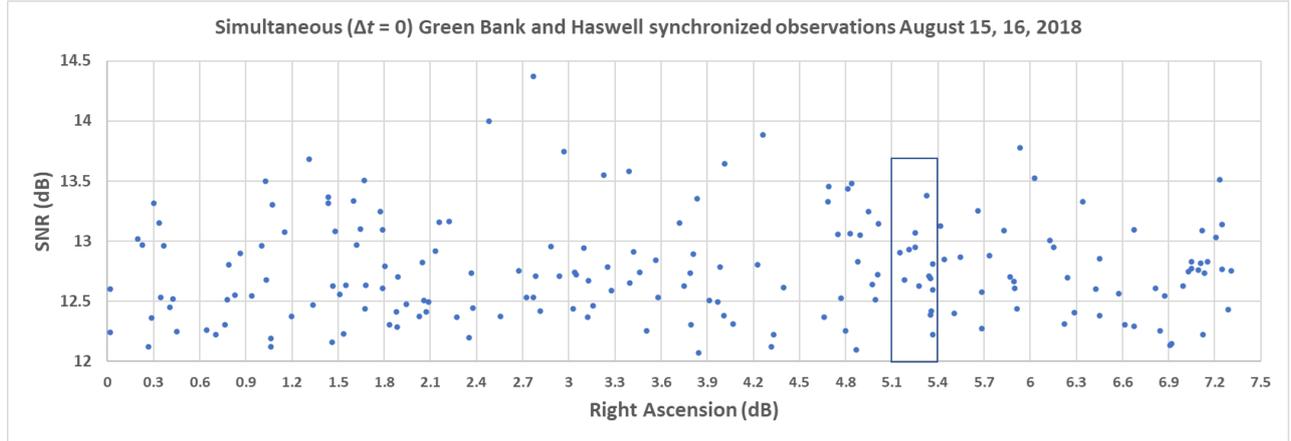

**Figure 4: SNR vs. *RA*** of pulse pairs during August 15, 16, 2018 observations. SNR is the higher of the SNRs of the Green Bank and Haswell SNRs having Δt = 0, | Δf | ≤ 400 Hz.

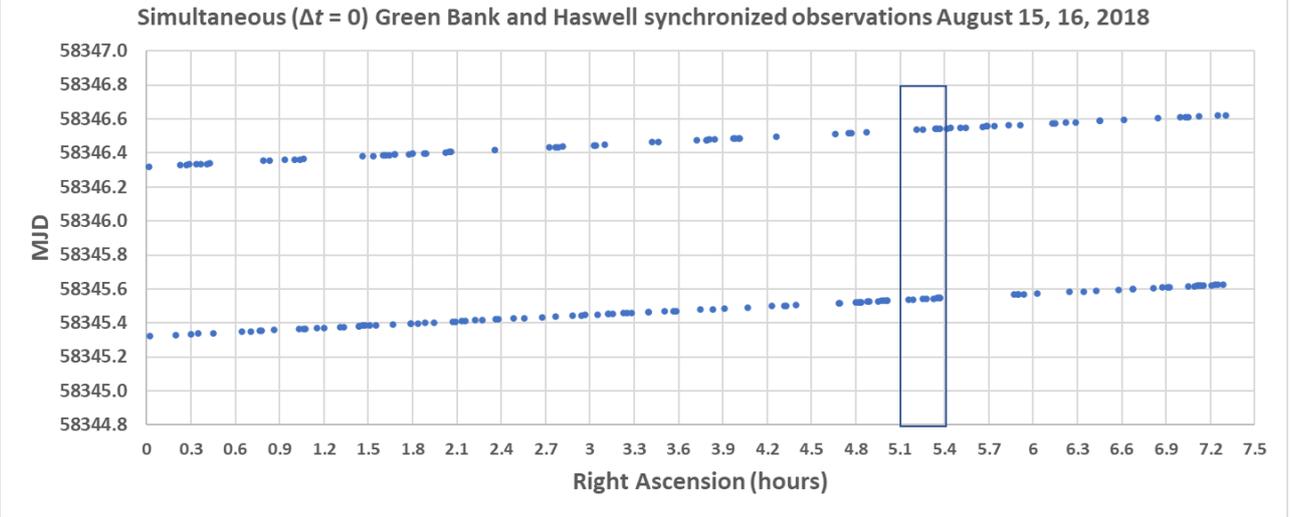

**Figure 5: *Modified Julian Date* vs. *RA*** of pulse pairs during August 15, 16, 2018 observations. SNR > 12.0 dB, Δt = 0, | Δf | ≤ 400 Hz.

| RA (hrs) | MJD GreenBank | MJD Haswell | SNR_MAX | Freq. Green Bank | Freq. Haswell | Δf (Hz) |
|---|---|---|---|---|---|---|
| **5.183775** | **58345.5380613** | **58345.5380613** | **13.293** | **1440.9286091** | **1440.9286091** | **0.0** |
| 7.028858 | 58346.6119994 | 58346.6119994 | 13.038 | 1436.4058451 | 1436.4058468 | -1.7 |
| **5.252898** | **58346.5382031** | **58346.5382031** | **12.495** | **1447.3290284** | **1447.3290277** | **0.7** |
| 23.14945 | 58345.2873177 | 58345.2873177 | 12.425 | 1426.2040372 | 1426.2040341 | 3.1 |
| 21.85269 | 58346.2307031 | 58346.2307031 | 12.312 | 1442.6352291 | 1442.6352262 | 2.9 |

**Table 1: Five highest SNRs (dB) of Green Bank and Haswell pulse pairs** having Δt = 0, |Δf| ≤ 3.1 Hz during August 15, 16, 2018 observations, are sorted by decreasing SNR_MAX. SNR_MAX is the higher value of the SNRs of the Green Bank and Haswell SNRs. Machine post-processing excised RFI. Haswell frequencies (MHz) are Earth rotation Doppler-compensated to Green Bank reference. Green Bank Forty Foot Telescope pointing azimuth is 180°. The likelihood of observing a second pulse pair, in *RA* range 5.183775 ± (5.252898-5.183775), in two tries, is approximately 0.017, given the measured *RA* range scanned during two days of observing with synchronized telescopes.

### D. August 16, 2018 observation indicates anomalies associated with $\Delta t = 0$, $|\Delta f| \leq 0.7$ Hz event at 5.1–5.4 hours RA, -7.6° DEC telescope pointing

**Figures 6–7** plot measurements relevant to SNR threshold crossings associated with $\Delta t = 0$, $|\Delta f| \leq 0.7$ Hz pulse pairs observed at Haswell and Green Bank. **Figure 6** indicates support for the AWGN model hypothesis while **Figure 7** implies that the AWGN model hypothesis is falsified, to the degree that the AWGN model posterior probability estimates a value of 0.014.

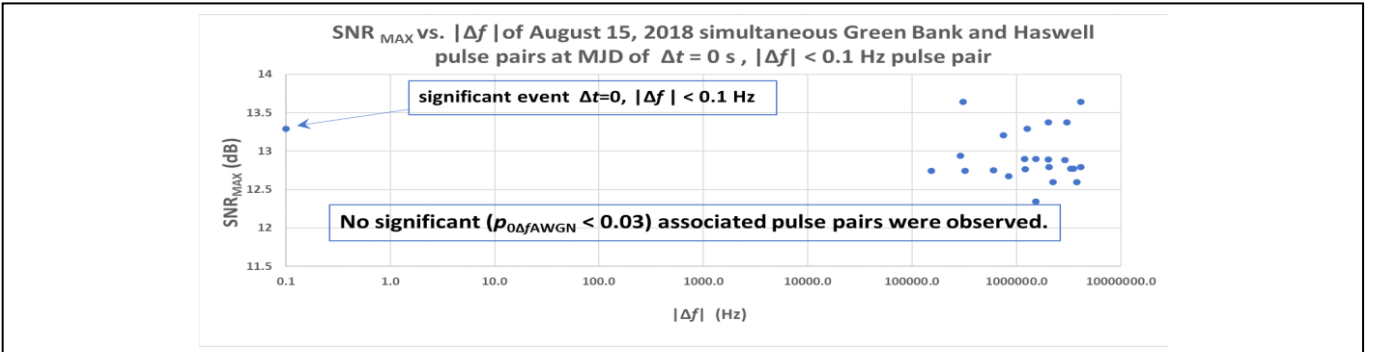

**Figure 6:** 5.183775 hours *RA* associated pulse pairs at Green Bank and Haswell, August 15, 2018, observed at the same MJD as a $\Delta f = 0$ Hz (plotted at 0.1 Hz) Green Bank and Haswell pair of pulses. No significant associated pulses were observed having $\Delta t = 0$ at the same MJD as the $\Delta f = 0$ Hz Green Bank and Haswell pulse pair. Bayesian inference indicates that the AWGN model hypothesis is supported by the absence of an associated pulse pair in the data.

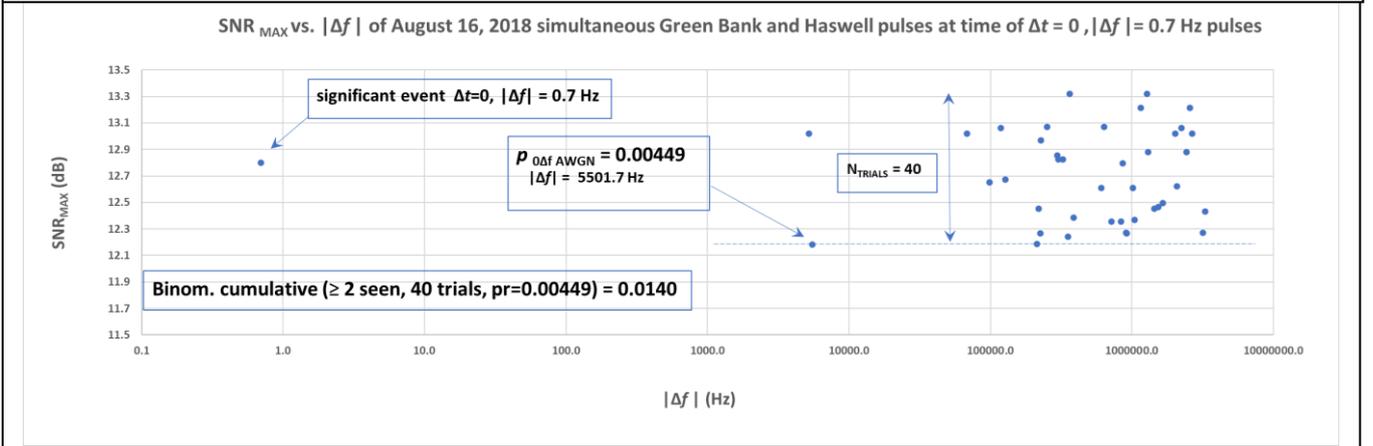

**Figure 7:** 5.252898 hours *RA* associated pulse pairs at Green Bank and Haswell, August 16, 2018, observed at the same MJD as a $|\Delta f| = 0.7$ Hz Green Bank and Haswell pair of pulses, indicate a binomial summed density, cumulative likelihood function value of 0.0140, given AWGN. The two associated pulse pairs, plotted at $|\Delta f| = 5501.7$ Hz and 5215.7 Hz, each presented one pulse of the pulse pair at Green Bank and the other pulse of the pulse pair at Haswell. Using Bayesian inference, assuming that the prior probability of the validity of the AWGN model is nearly one, and the probability of valid data is nearly one, and given the **Appendix C, Method A** binomial cumulative likelihood at 0.0140, the posterior probability of the AWGN model explaining data up to and including August 16, 2018 data is approximately 0.0140.

### E. April 2, 3, 2019 observations present two $\Delta t = 0$, $|\Delta f| \leq 15.5$ Hz anomalies at 5.1–5.4 hours RA, -7.6° DEC telescope pointing

**Table 2** details $\Delta t = 0$, $\Delta f \approx 0$ Hz pulse pairs recorded over *RA* ranging from 0 to 24 hours, SNR $\geq 12.0$ dB, for the April 2, 3, 2019 observation run. **Figures 8–13** detail simultaneous and associated pulse pairs for the April 2,3, 2019 observation run. The likelihood function used in **Figures 12–13** is described in **Appendix C, Method A**.

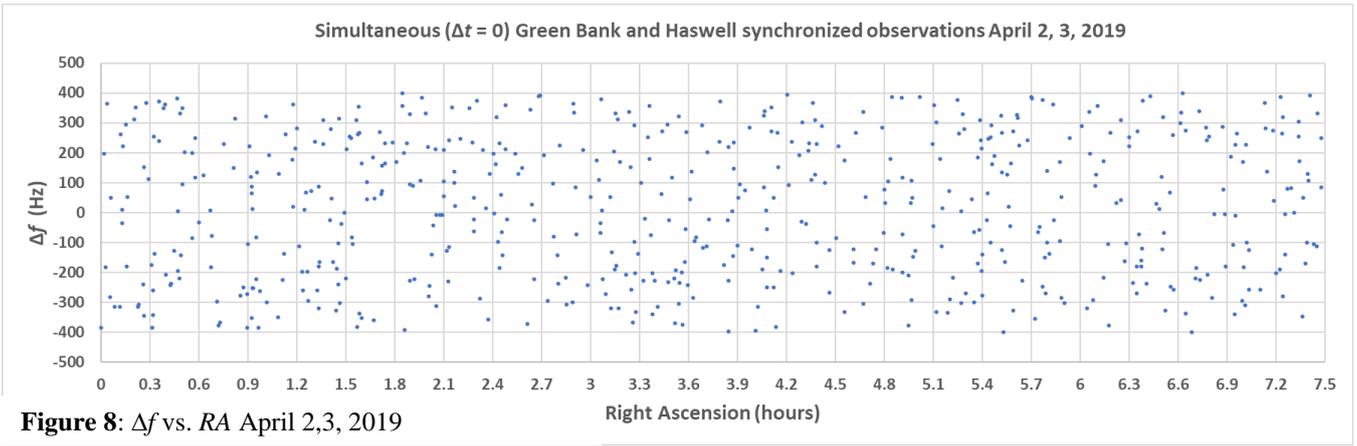

**Figure 8**: Δ*f* vs. *RA* April 2,3, 2019

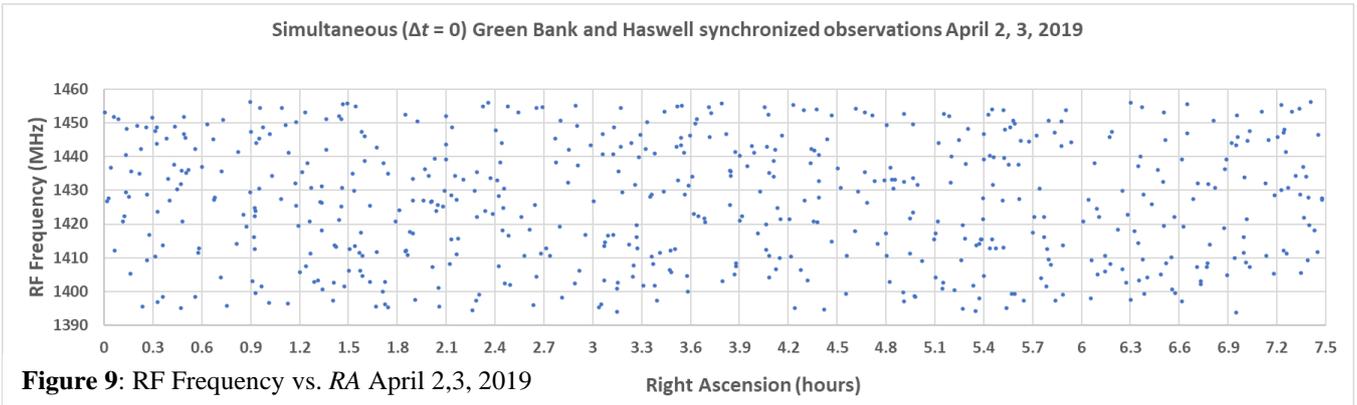

**Figure 9**: RF Frequency vs. *RA* April 2,3, 2019

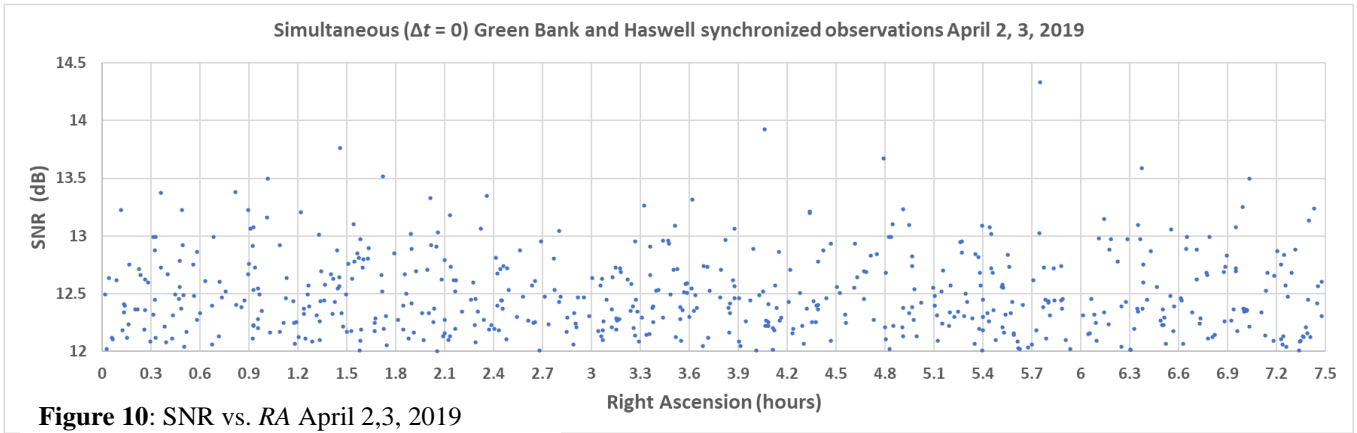

**Figure 10**: SNR vs. *RA* April 2,3, 2019

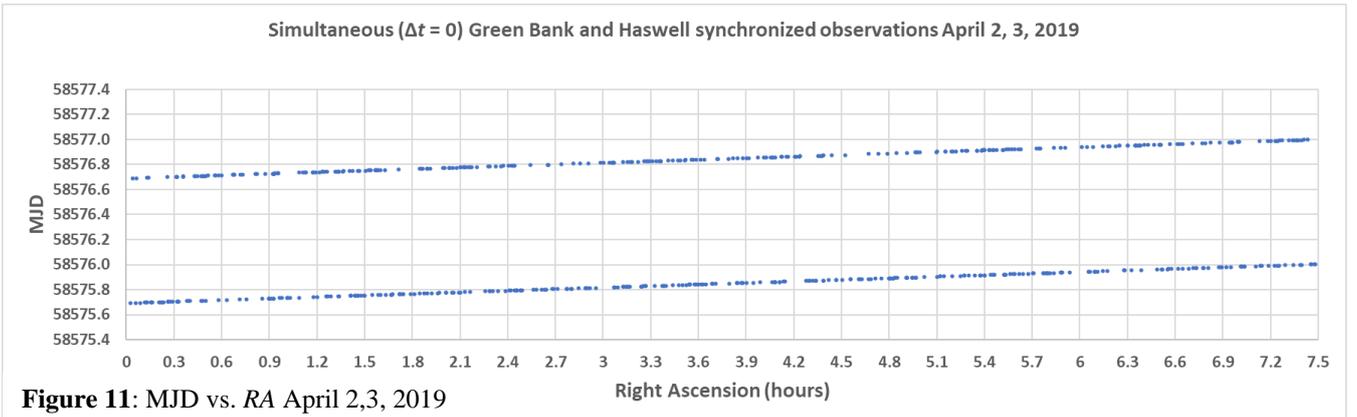

**Figure 11**: MJD vs. *RA* April 2,3, 2019

| RA (hrs) | MJD GreenBank | MJD Haswell | SNR_MAX | Freq. Green Bank | Freq. Haswell | Δf (Hz) |
|---|---|---|---|---|---|---|
| 2.358213 | 58576.7871904 | 58576.7871904 | 13.346 | 1455.9449650 | 1455.9449508 | 14.2 |
| 9.287270 | 58577.0751128 | 58577.0751128 | 13.289 | 1417.4416430 | 1417.4416352 | 7.8 |
| 12.258186 | 58576.2012934 | 58576.2012934 | 13.158 | 1421.3716433 | 1421.3716373 | 6.1 |
| 23.412879 | 58576.6648032 | 58576.6648032 | 12.919 | 1406.8068265 | 1406.8068143 | 12.2 |
| **5.270635** | **58575.9109404** | **58575.9109404** | **12.854** | **1395.0271667** | **1395.0271609** | **5.8** |
| 23.830244 | 58577.6794155 | 58577.6794155 | 12.798 | 1399.1676383 | 1399.1676267 | 11.6 |
| **5.156062** | **58576.9034491** | **58576.9034491** | **12.698** | **1452.7422110** | **1452.7421955** | **15.5** |

**Table 2: Seven highest SNR (dB) pulse pairs during April 2, 3, 2019 observation run**, having SNR ≥ 12.698 dB, Δt =0, 5.8 Hz ≤ Δf ≤ 15.5 Hz. An apparent significance of the 5.1–5.4 hours *RA* region appears, after reviewing **Table 1** second-event likelihood ≈ 0.017, while considering **Table 2**, indicating two or more events in seven tries, at event probability = 0.3 hours / 24 hours, yielding an AWGN model binomial cumulative likelihood ≈ 0.0031.

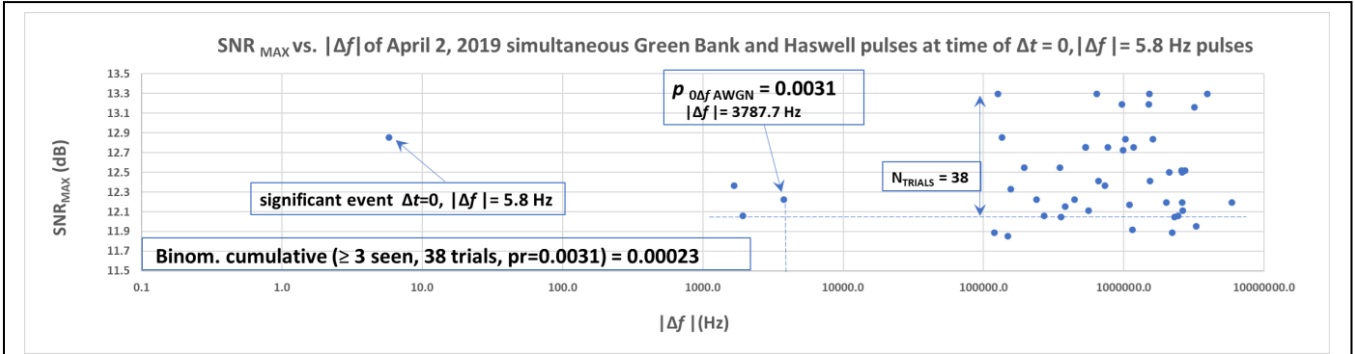

**Figure 12:** **5.270635 hours *RA*** associated pulse pairs at Green Bank and Haswell, observed at the same time as a |Δf| = 5.8 Hz Green Bank and Haswell pair of pulses, indicate a binomial cumulative likelihood function value of 0.00023, given an AWGN model. The lowest |Δf| associated pulse pair, among the three pulse pairs, was recorded at Haswell, while the next two higher |Δf| pulse pairs presented one pulse element at Green Bank and the other at Haswell.

Using Bayesian inference, assuming August 16, 2018 prior probability of the validity of the AWGN model, assuming the probability of valid data at nearly one, and given the binomial cumulative likelihood at 0.00023, the posterior probability of the AWGN model explaining data as of April 2, 2019, is approximately 0.00023 multiplied by the post-August 16, 2018 AWGN prior model probability at 0.0140, equal to 3.2 x 10⁻⁶.

One of the three associated pulse pairs, at |Δf| = 1661.4 Hz, measured a suspected-RFI RF frequency of 1409.021486 MHz. If this pulse pair is assumed to be 1 MHz harmonically related RFI, the binomial cumulative likelihood function increases from 0.00023 to 0.0060, and the April 2, 2019 posterior increases to 8.5 x 10⁻⁵.

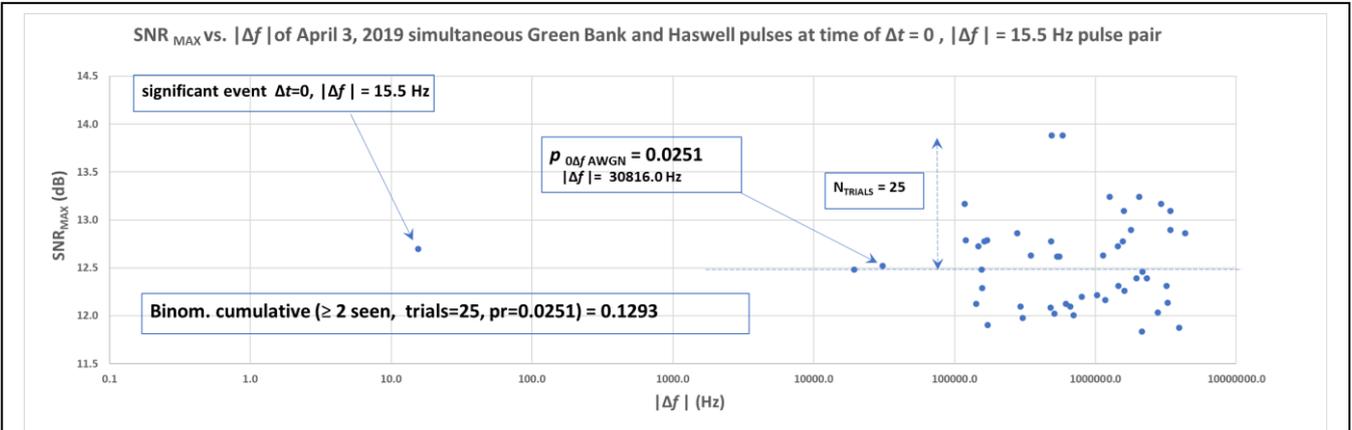

**Figure 13:** **5.156062 hours *RA*** associated pulse pairs at Green Bank and Haswell, observed at the same time as a |Δf| =15.5 Hz Green Bank and Haswell pair of pulses, indicate a binomial cumulative likelihood function value of 0.1293, given an AWGN model. The two associated pulse pairs presented each element of the pulse pair at Green Bank and Haswell. The posterior probability of the AWGN model explaining the observed data of April 3, 2019, is slightly reduced from the prior AWGN model probability, determined by the posterior of April 2, 2019, 8.5 x 10⁻⁵ multiplied by 0.1293 = 1.1 x 10⁻⁵.

### F. 164 beam transits of the Twenty-six Foot telescope

A beam transit is defined to be the telescope FWHM antenna response transiting an *RA* at a *DEC* of -7.6°, once per day, due to Earth's rotation.

Machine post-processing was utilized to examine 164 beam transits, during 181 days of Twenty-six Foot telescope observations. The beam transit test was conducted between September 4, 2020 and March 4, 2021.

The number of beam transits is less than the number of elapsed observation days due to exclusion of raw data files, caused by observation outage, suspected short-term RFI described below, and calibration that occurs shortly after new raw data files are created, potentially corrupting the 5.1 to 5.4 hour *RA* region of interest.

Three days of raw data files were manually excised from machine post-processing, after discovering bursts of suspected RFI pulses in the range of 4.318 – 4.319 hours *RA* on MJD 59259, 9.158 – 9.196 hours *RA* on MJD 59275, and 5.26 – 5.27 hours *RA* on MJD 59270. As these three sets of pulses did not exhibit the properties expected of AWGN, these three days were not included in the files scanned in machine post-processing.

**Figure 14** plots MJD vs *RA* of $\Delta t = 0$, $|\Delta f| \leq 400$ Hz orthogonal circular polarized pulse pair events, from the machine post-processed output file, at SNR ≥ 13.0 dB.

The MJD vs. *RA* is shown in **Figure 15** for an SNR ≥ 13.532 dB.

**Figure 16** plots SNR vs. *RA*. **Figures 17–18** plot RF Frequency and $\Delta f$ for SNR ≥ 13.532 dB. SNR is the higher of the two SNR values in a $\Delta t = 0$, $|\Delta f| \leq 400$ Hz orthogonal polarized pulse pair.

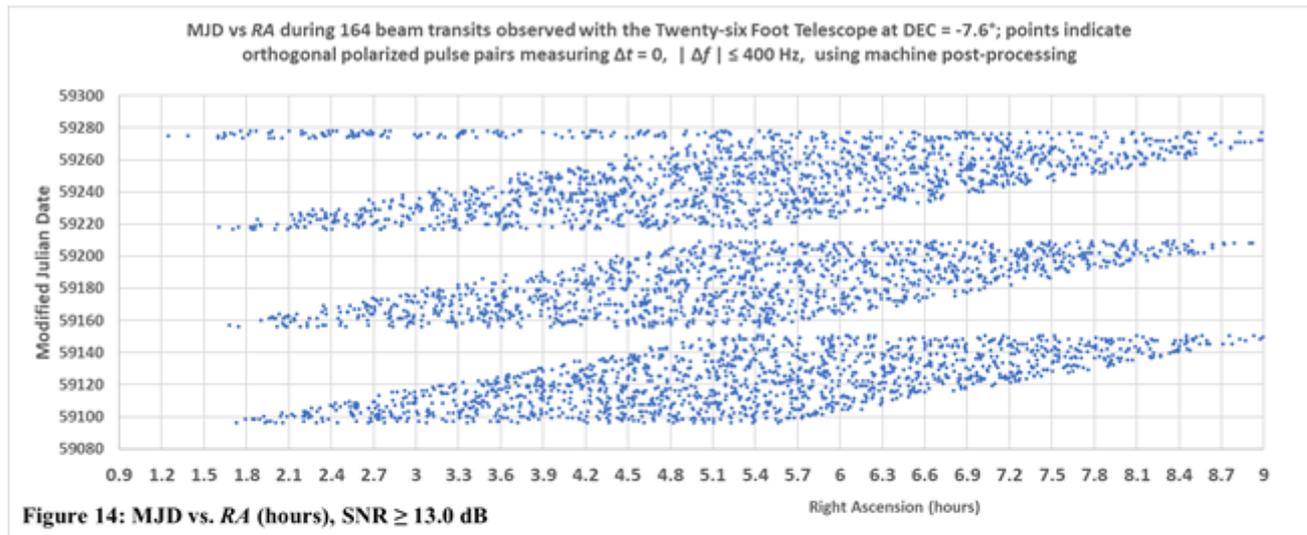

**Figure 14: MJD vs. *RA* (hours), SNR ≥ 13.0 dB**

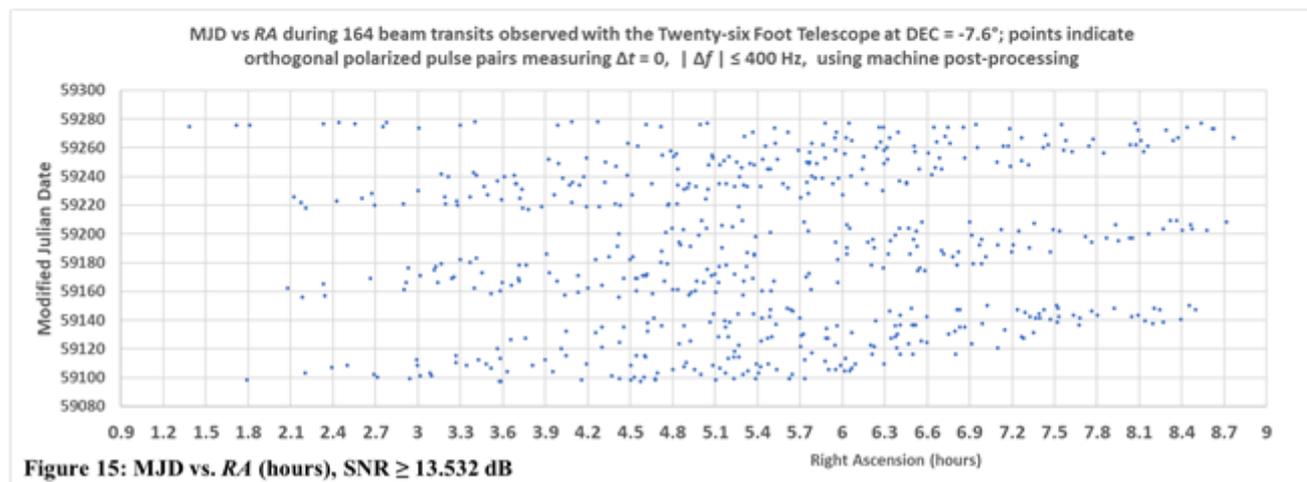

**Figure 15: MJD vs. *RA* (hours), SNR ≥ 13.532 dB**

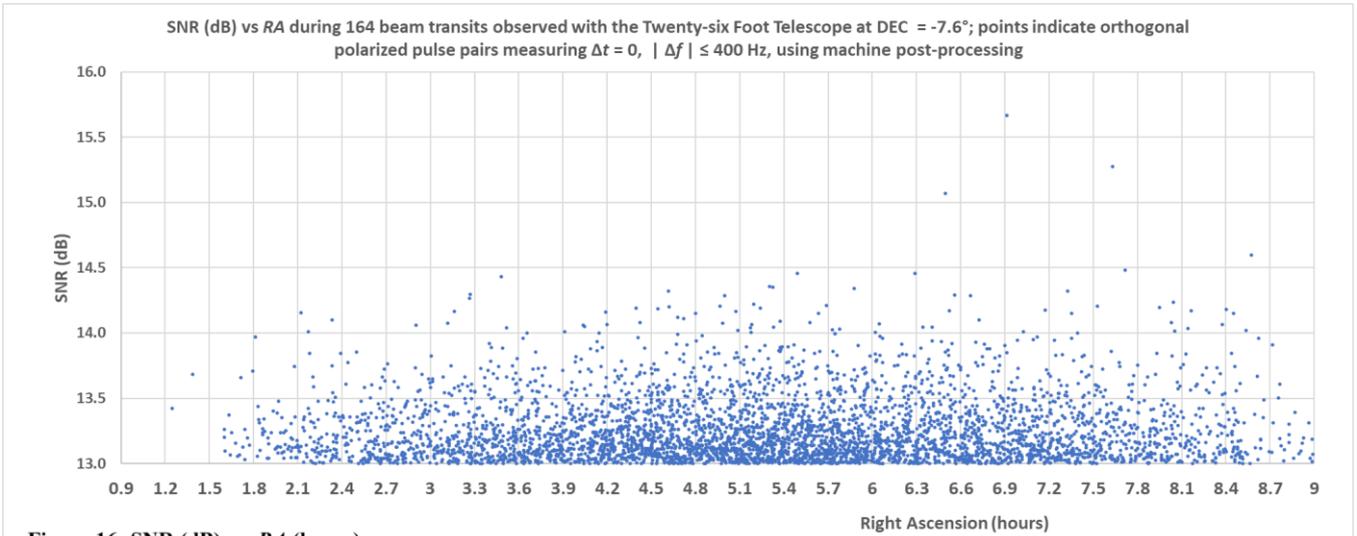

**Figure 16: SNR (dB) vs. *RA* (hours)**

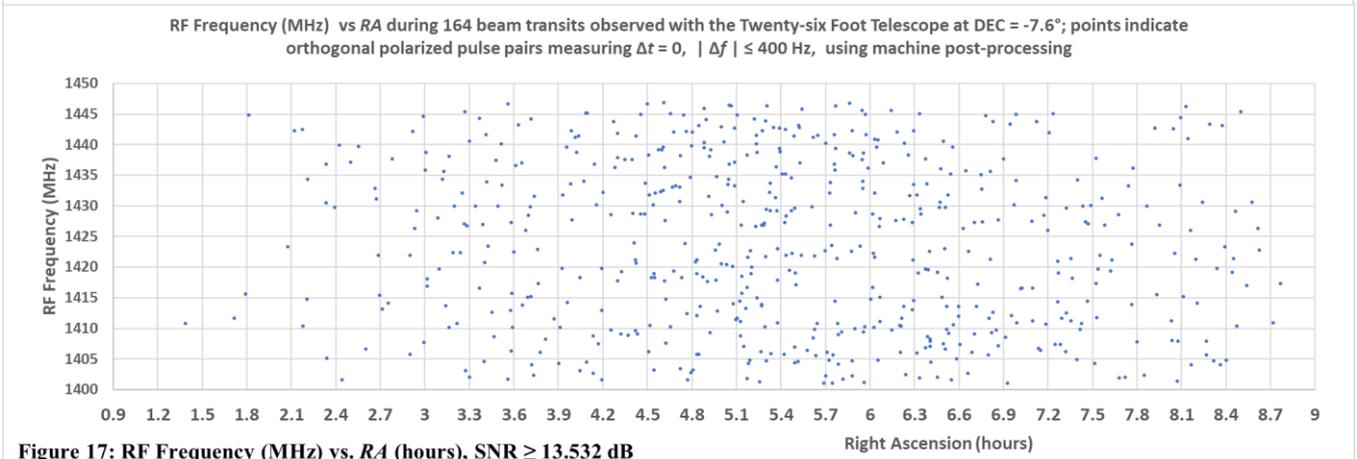

**Figure 17: RF Frequency (MHz) vs. *RA* (hours), SNR ≥ 13.532 dB**

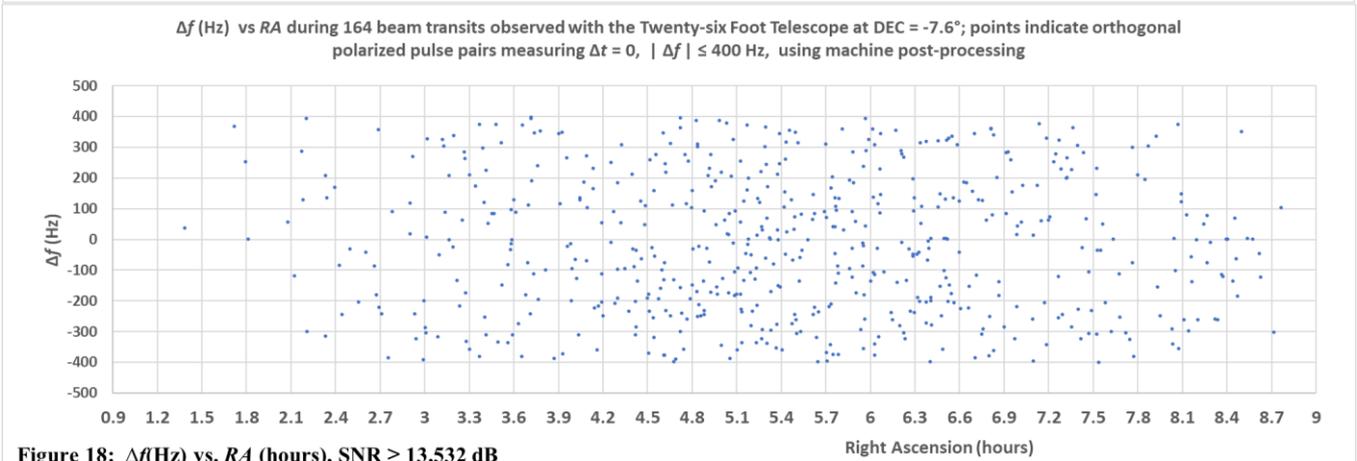

**Figure 18: Δ*f*(Hz) vs. *RA* (hours), SNR ≥ 13.532 dB**

**Figure 19** plots the SNR-sorted binomial probability density values of $\Delta t = 0$, $|\Delta f| \leq 400$ Hz orthogonal circular polarized pulse pairs, observed in five *RA* ranges, covering 4.5 to 6.0 hours *RA*, during 164 beam transits, spanning 181 days. The binomial probability density calculates a likelihood function, based on AWGN-caused probability, i.e. that a count of $\Delta t = 0$, $|\Delta f| \leq 400$ Hz events will be seen above an SNR threshold level identified by the SNR-ordered trials number, within an *RA* range. The likelihood function is described in **APPENDIX C** *AWGN-model likelihood of $\Delta t$ $\Delta f$ elements in hypothesized energy bursts*, **Method B**.

In the binomial calculation, the probability of a pulse pair event occurring within each *RA* range is estimated by dividing the number of events counted in each *RA* range by the number of total events, based on the assumption that

AWGN is the cause of the events. The event probabilities measured in the five *RA* ranges, in **Figures 19–20**, are 0.065, 0.067, 0.078, 0.072, and 0.067.

AWGN theory predicts a central *RA* event probability of $0.075 = 0.3$ RA hours / 4.0 MJD hours, assuming a uniform density of events, and predicts linearly decreasing event probability as the *RA* differs from the central *RA* value.

**Figure 20** expands the plot in **Figure 19** to display binomial likelihood *L* values for the highest 1000 SNR threshold crossings of $\Delta t = 0$, $|\Delta f| \leq 400$ Hz pulse pairs. Low *L* values indicate either anomalous presence, or anomalous absence of pulse pairs. Decreasing discontinuities indicate anomalous presence, while increasing discontinuities indicate anomalous absence of pulse pairs. The decreasing discontinuities evident in the 5.1–5.4 hour *RA* range indicate an anomalous presence of orthogonal polarization $\Delta t = 0$, $|\Delta f| \leq 400$ Hz pulse pairs. The double minima in 5.1 –5.4 hours *RA* calculate a composite normalized AWGN likelihood $L \approx 0.017$, and a posterior approximately 0.017 times the prior AWGN model validity, post April 3, 2019.

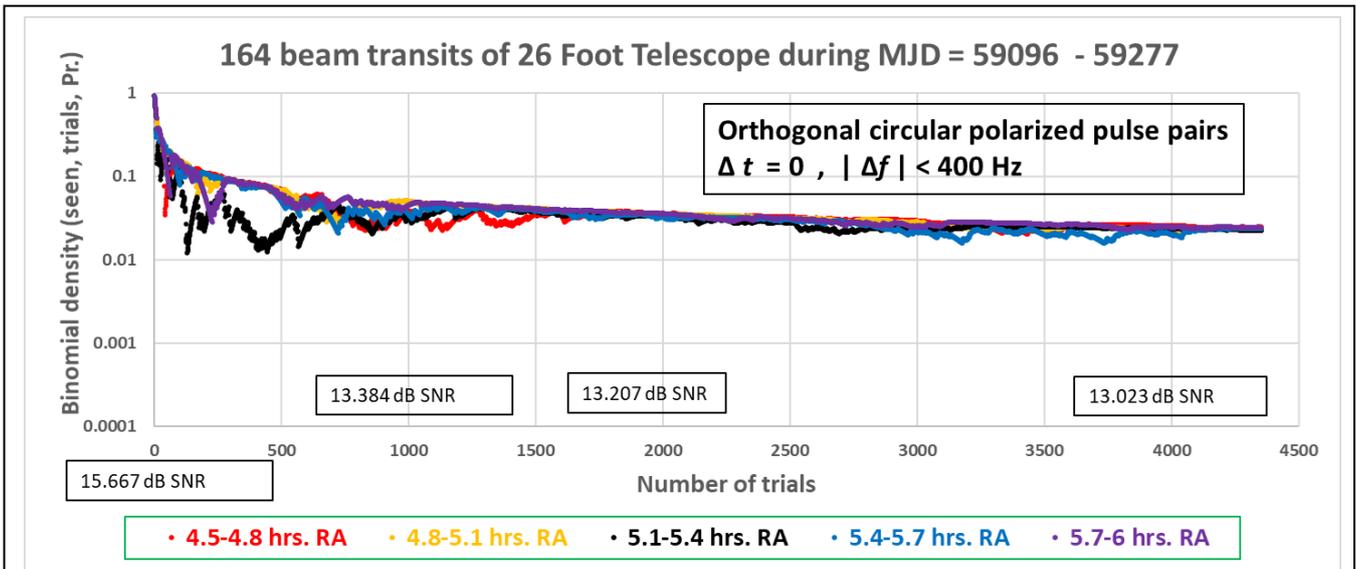

**Figure 19: 5.1–5.4 hrs *RA*, and four adjacent/alternate *RA* windows (4.5–5.1 hrs and 5.4–6.0 hrs *RA*), binomial density likelihood function, *L* vs. Number of trials**.

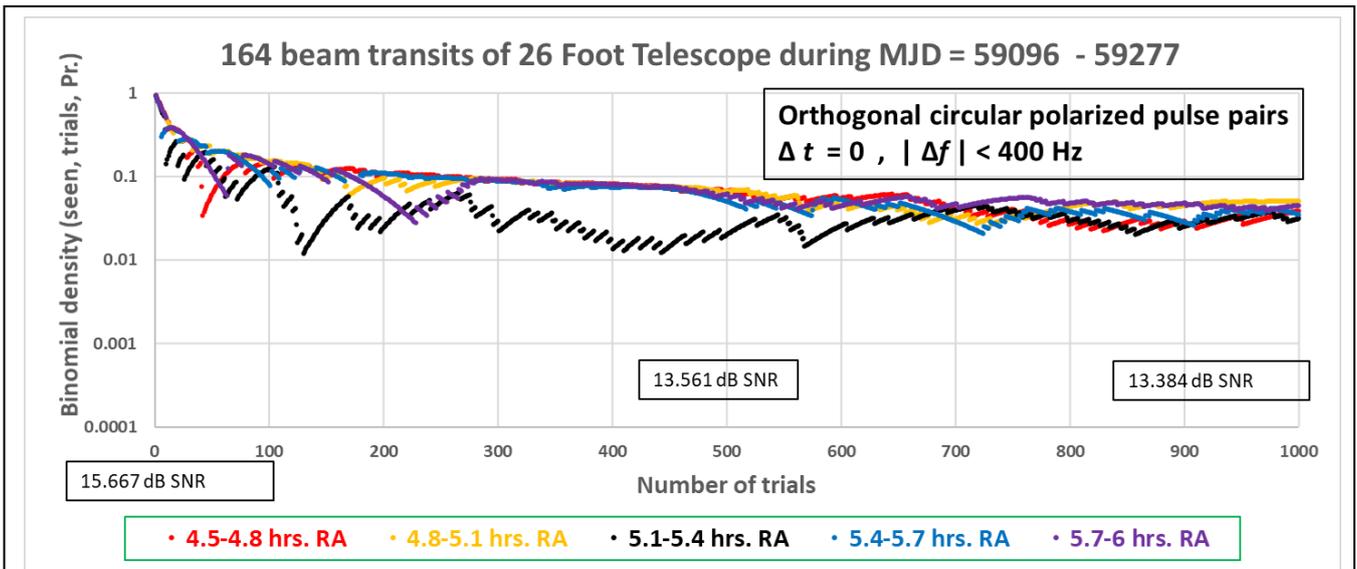

**Figure 20:** (Expansion of **Figure 19**, highest 1000 SNR trials) **5.1–5.4 hrs *RA*, and four adjacent/alternate *RA* windows (4.5–5.1 hrs and 5.4–6.0 hrs *RA*), binomial density likelihood function, *L* vs. Number of trials**. Decreasing discontinuities indicate an anomalous presence of pulse pairs.

**Figure 21** plots machine post-processing output of 44 days of the Twenty-six Foot Telescope raw data, with an artificial AWGN noise source connected to one polarization channel of the receiver, while using the machine hyperparameters as in **Figures 19 – 20.** The shape of the 5.1–5.4 hours *RA* low *L* values indicates an anomalous absence of pulse pairs, due to the increasing binomial value in the discontinuities.

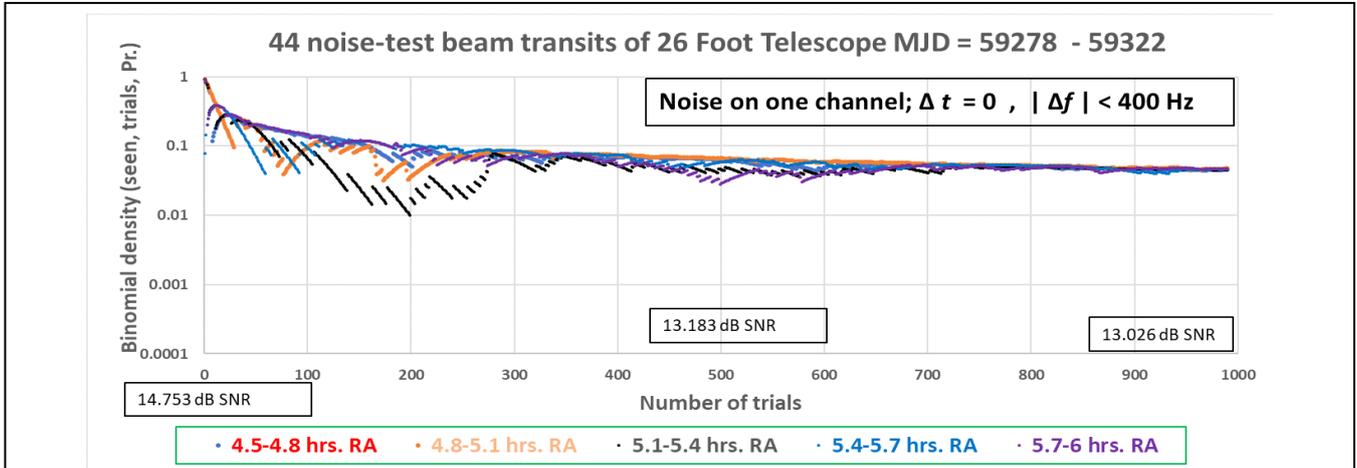

**Figure 21: Noise test, using machine post-processing system:** 5.1–5.4 hrs *RA*, and four adjacent/alternate *RA* windows binomial density likelihood function *L*, vs. Number of trials, up to 990 trials, with artificial AWGN noise source input to one polarization channel. The 4.8–5.1 hour *RA* range indicates decreasing discontinuities at $L \approx 0.03$, possibly due to the 44 day, reduced sample population size, a topic of investigation in equipment-cause hypotheses.

**Figure 22** plots the result of manual adjustment of the *RA* range and center value, to 0.22 hours *RA* range, at 5.28 hours center *RA*, and presents a calculated minimum AWGN likelihood $L = 0.00144$, at 130 trials, approximately 100 times less likely than expected in AWGN. The **Figure 22** plot uses event probability values across five *RA* ranges = 0.046, 0.054, 0.055, 0.055, and 0.050.

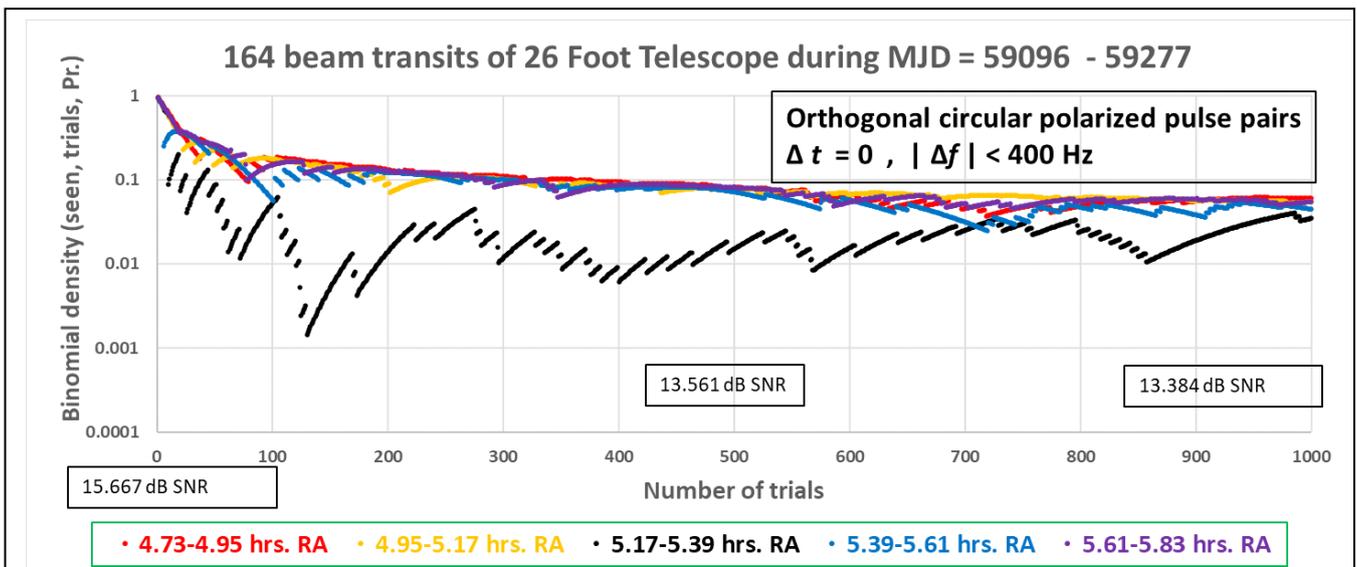

**Figure 22:** *RA* range and center *RA* adjusted: 5.17–5.39 hours; *RA* and four adjacent/alternate *RA* windows binomial density likelihood function *L*, vs. Number of trials, up to 1000 trials. The center *RA* and the *RA* range were manually adjusted in a search for lowest composite binomial density. Likelihood *L*, at 130 trials, calculates a value approximately 100 times lower than expected in an AWGN-explained model. The adjusted *RA* range will be used to calculate posterior probabilities, testing an *RA*-updated hypothesis in a currently running beam transit experiment.

## VI. Discussion

Conclusions regarding anomalies observed in this experimental work are problematic, for at least six reasons.

1. **Auxiliary hypotheses** – The AWGN hypothesis, as proposed, appears to be falsified, based on posterior probability values and Bayesian inference, resulting from observations and machine post-processing. However, there are many auxiliary hypotheses that may be developed, modifying the proposed AWGN hypothesis, and explaining that a modified AWGN hypothesis is indeed a likely explanation for the observed anomalies. An example of a reasonable auxiliary hypothesis is that the machine RFI excision processes are imperfect and are introducing RFI-induced $\Delta t\,\Delta f$ events into post-processed results. The apparent *RA* dependency in the 164 beam transit test might be explained with other auxiliary hypotheses, e.g. celestially correlated unexcised RFI, corrupting the AWGN model, and/or data population effects.

2. **The number of anomalous pulse pairs** – The three observation runs present a total of approximately twenty $\Delta t\,\Delta f$ pulse pairs indicating low AWGN likelihood, reported by one observer. Many auxiliary and alternate hypotheses likely explain the anomalies, e.g. RFI and receiving equipment.

3. **Bias due to prior knowledge of some experimental results** – Details of the AWGN hypothesis in this experiment were developed after the August 15, 16, 2018 observation, and refined after the April 2, 3, 2019 observation, which were manually post-processed, and publicly reported. A plausible argument exists that this work's hypothesis was developed based on prior knowledge of experimental results. Given strength in this argument, only the 164 beam transit test might be acceptable as a valid experimental test of the current hypothesis. A counter-argument to the a-priori knowledge argument may be made as follows. An informal hypothesis has been present in the investigative, pre-formal-hypothesis work, prior to 2018. This informal hypothesis stated that receivers that search for interstellar signals should utilize mechanisms designed to optimally receive signals from transmitters designed for high energy efficiency and high channel capacity, and that an AWGN hypothesis may be falsified by building and operating receivers optimized to receive hypothesized transmissions. In this sense, the primary addition to the informal hypothesis, post-August 15, 16, 2018, is the predicted significance of the 5.25 hour *RA*, added to the -7.6° *DEC* celestial direction. The April 2, 3, 2019 observations and the 164 beam transit test, each may then be considered to be tests of a post-August 15, 16, 2018 hypothesis, together with its *RA* and *DEC* prediction. In this scenario, the probability that the August 15, 16, 2018 measurements are explained by AWGN becomes the prior AWGN model probability, applied to the observations subsequent to August 15, 16, 2018, to obtain posterior probabilities, using Bayesian inference.

4. **Isolated falsification** – Falsification of one hypothesis to explain data anomalies, while many other valid auxiliary and alternate hypotheses are untested, generally leads to conclusions regarding further work, and generally does not lead to meaningful conclusions that explain anomalous observations.

5. **A subset of observation runs was machine post-processed** – The full set of observations in 2017–2021 was not machine post-processed in this work. An argument ensues that the posterior probabilities calculated in this experiment should be increased by a trial count factor, since machine post-processing of other observations might show strong support of the AWGN hypothesis. Further, the processing of all observation runs may reveal an equipment cause of the AWGN anomalies of August 15, 16, 2018, April 2, 3, 2019, and 164 beam transit tests.

6. **Assumptions are present** – In the calculation of Bayesian posteriors, measurement data is assumed to be equally valid, at nearly one, across observation runs. RFI and human-made communication systems are confounding issues in data validity. Invalid data in the test of an AWGN hypothesis leads to posteriors that support the AWGN hypothesis.

## VII. Conclusions

Considering the observations, calculated Bayesian posteriors, and arguments, a conclusion is suggested that the AWGN model developed in this experimental hypothesis is presently not supported by the findings of this experiment, albeit for unknown reasons. Conclusions are generally incomplete absent an underlying explanation. Apparent falsification of the AWGN hypothesis compels further work.

## VIII. Further work

1. Another long term Twenty-six Foot Telescope beam transit test is being conducted at -7.6° *DEC*. A modified AWGN hypothesis using a 5.17–5.39 hour *RA* range is being tested. Another polarization channel with back-end processing is under construction, to provide a total of three pairs of $\Delta t\,\Delta f$ measurements.

2. Interferometer and phased array antennas are under construction, to improve the measurement precision of the arrival direction of $\Delta t\,\Delta f$ pulse pairs. The RF frequencies of the observed pulses associated with $\Delta t\,\Delta f$ anomalies do not appear concentrated, from one $\Delta t\,\Delta f$ anomaly to the next. The appearance of wide bandwidth elements in a received signal leads to an idea that phased array elements may be used to estimate AOA, with reduced measurement ambiguity caused by phase wrapping and grating lobes inherent in sparsely filled antenna arrays. Precise pointing information and analysis of $\Delta t\,\Delta f$ anomalies improve the effectiveness of follow-up observations.

3. Studying the three days of manual RFI excisions in the 164 beam transit test, might help determine if the machine post-process algorithm should, or should not, have found pulse pairs as suspected RFI.

4. Continuously operating local RFI receivers are planned, having receiver sensitivity similar to telescope receivers, and connected to antennas having various beamwidths, polarizations, locations and pointing directions.

5. Additional long term artificial noise source tests are planned to test spurious signal hypotheses, and analyze post processing results when artificial noise is applied to one or more polarization channels.

6. Enhance post-processing of anomalous observations in a search for significant $\Delta t\, \Delta f$ anomalies, including associated pulses, and anomalies at $|\Delta t| > 0$.

7. Continue synchronized and single telescope observations, and measure the repeatability of the observed $RA\ DEC$ anomaly.

8. Machine post-process archived observations.

9. Design and implement metrics of correlation of observed candidate $\Delta t\ \Delta f$ anomalies with excised RFI.

10. Augment receiver systems to down-convert the 1660–1670 MHz and 2690–2700 MHz radio astronomy bands to digitizer frequency range.

11. Install additional receiver polarization channels, to a total number $N_{POL} \approx 6$, to receive and process $N_{POL}(N_{POL}-1)/2 = 15$ channels of $\Delta t\, \Delta f$ pulse pairs.

12. Develop and test alternate and auxiliary hypotheses.

13. Seek means of corroboration.

## IX.   ACKNOWLEDGMENTS

Special thanks are given to Steve Plock and Ed Corn of the Deep Space Exploration Society, for their many valuable contributions to the systems, operations and observations of the Plishner Telescope, to the Deep Space Exploration Society members who provided their valuable expertise and work, and to the workers of the SETI Institute, Breakthrough Listen and the U.C. Berkeley SETI Research Center, for ideas and helpful guidance. Special thanks are given to the workers of the Green Bank Observatory, all of whom helped make the Forty Foot Telescope observations possible. Special thanks are given to family and friends for helpful advice.

# APPENDIXES

## APPENDIX A
### *AWGN-model likelihood of polarization*

Polarization is expected to be uniform on the Poincaré sphere, due to the polarization independent properties of AWGN.

The choice of receiver polarization values is a component within the overall design of a receiver's optimal filters, matched to discover and decode a transmitted signal. Two $\Delta t\, \Delta f$ signal discovery receivers, having differing polarized antennas, sample the Poincaré sphere at two points, each providing a filter matched to a transmitter quantized polarization value, modified by dynamic geometry and propagation.

If transmitted $\Delta t\, \Delta f$ discovery signals fill the Poincaré sphere, with quantization, then $\Delta t\, \Delta f$ signals are discoverable using single-polarized receivers, of any polarization, albeit at reduced rate, due to increased transmit-receive polarization mismatch. Multiple polarizations may be sampled to improve the $\Delta t\, \Delta f$ signal discovery rate.

## APPENDIX B
### AWGN-model likelihood of SNR

Ricean statistics describe the amplitude probability density of a sinusoidal signal combined with independent in-phase and quadrature components of Gaussian noise [15]. Given that only noise is present in a received signal, Ricean statistics reduce to Rayleigh statistics. The ratio of the probability density functions of Ricean to Rayleigh statistics, as a function of an SNR crossing threshold, may be used to estimate the probability that a sinusoidal signal combined with AWGN will result in an SNR threshold crossing, relative to the probability that AWGN alone will result in the same SNR threshold crossing. The calculation of this probability follows.

Ricean amplitude probability has the following density [15][16],

$$p_{RI}(r) = (r/\sigma^2) \exp(-(r^2 + s^2)/2\sigma^2) \, I_0(r \, s \, /\sigma^2) \qquad (7)$$

where r is the amplitude of the composite sinusoidal signal and Gaussian noise, $\sigma^2$ is the variance of the Gaussian noise, s is the amplitude of the sinusoidal component and $I_0(\bullet)$ is the zeroth-order modified Bessel function of the first kind.

Rayleigh amplitude statistics reduce (7) to

$$p_{RA}(r) = (r/\sigma^2) \exp(-r^2/2\sigma^2) \qquad (8)$$

The Ricean to Rayleigh density ratio is

$$p_{RI}(r) \, / \, p_{RA}(r) = \exp(-s^2/2\sigma^2) \, I_0(r \, s \, /\sigma^2). \qquad (9)$$

The effect of the density ratio may be estimated by examining the effect of increased sinusoidal signal amplitude, at an amplitude detection threshold of 5 times ($\approx$14 dB) the average amplitude of noise. The Ricean to Rayleigh probability density ratio increases from approximately 17 to 14,612 after the sinusoidal signal amplitude is increased from a value equal to the average noise amplitude, to a value four times the average noise amplitude.

The effect of the increase in (9) implies that a reception process that examines high SNR pulse pair measurements, among $\Delta t \, \Delta f$ anomalies, provides an increased likelihood that intentionally enhanced amplitude $\Delta t \, \Delta f$ discovery signals will be discovered by a receiver. The increase of the density function ratio, at an SNR threshold, points to what seems to be natural shared knowledge between transmitter and receiver, i.e. knowledge that increasing the amplitude of particularly infrequent elements of a signal is useful to the discovery of a wideband energy-efficient transmitted signal.

The properties of the Ricean to Rayleigh density ratio point to a strong means of AWGN hypothesis falsification, since SNR threshold crossings may be used to seek anomalies in AWGN.

In the current work, the AWGN-model likelihood of SNR is applied in setting SNR thresholds for data capture, the excision of persistent and dynamic RFI, and the high-to-low SNR sorting of $\Delta t \, \Delta f$ pulse pairs, to calculate their significance in AWGN, based on other measurements of the signals, e.g. $RA$.

## APPENDIX C
### AWGN-model likelihood of $\Delta t$ $\Delta f$ discovery elements in hypothesized energy bursts

Assuming that Poisson point process statistics apply to high SNR events in AWGN, the expected probability of the presence of high SNR $\Delta t$ $\Delta f$ elements within AWGN may be estimated as follows.

The probability that zero Poisson distributed events will occur during a time $t$ is quantified by

$$p_0(t) = \exp(-rt) \qquad (10)$$

where $r$ is the average rate of events [3]. Equation (10) may be solved at a $t$ value giving $p_0(\Delta t_{50\%}) = 0.5$, where $\Delta t_{50\%}$ is defined as the median interarrival time, among $\Delta t > 0$, $\Delta f > 0$ signal elements, above an SNR level, to obtain a value of

$$r = r_{\Delta t} = \ln(2) \, / \, \Delta t_{50\%}. \qquad (11)$$

A similar $\Delta f$ rate parameter may be calculated, yielding $r_{\Delta f} = \ln(2) \, / \, \Delta f_{50\%}$. Using the two $r_{(\bullet)}$ values, the probability of a non-zero number of $\Delta t$ $\Delta f$ events, assuming independence of $\Delta t$ and $\Delta f$ events, is calculated using

$$p_{\Delta t \Delta f \, AWGN} (\Delta t, \Delta f) = (1 - \exp(-\ln(2) \, \Delta t/\Delta t_{50\%} \, )) \cdot (1 - \exp(-\ln(2) \, \Delta f/\Delta f_{50\%} \, )) \qquad (12)$$

and, at $\Delta t << \Delta t_{50\%}$ and $\Delta f << \Delta f_{50\%}$, approximated to

$$p_{\Delta t \Delta f \, AWGN} (\Delta t, \Delta f) \approx (\ln(2))^2 \, \Delta t \, \Delta f \, / \, \Delta t_{50\%} \Delta f_{50\%}. \qquad (13)$$

In measurements where $\Delta t = 0$,

$$p_{0\Delta f \, AWGN} (\Delta f) = \ln(2) \, \Delta f \, / \, \Delta f_{50\%} \qquad (14)$$

is calculated, where the 0 subscript in $p_{0\Delta f \, AWGN} (\Delta f)$ indicates that the trial population comprises $\Delta t = 0$ events.

The joint probability calculation in (13) may be used to estimate the event likelihood of the observation of a pulse pair having $\Delta t \, \Delta f$ elements, assuming an AWGN source model. The bivariate Poisson-based probability $p_{\Delta t \Delta f AWGN}(\Delta t, \Delta f)$ of the $\Delta t$ and $\Delta f$ random variables, improves the effectiveness of the discovery signal detection mechanism, compared to single variable probabilities.

The calculation of likelihood in this experiment is affected by the minimum values of $\Delta t = 0.27$ seconds, and $\Delta f = 3.7$ Hz, and the quantization of these two measurements. The presentation of results in this work involve $\Delta t = 0$ events, and the effect of quantization of $\Delta f$ is not considered.

Equation (14) is used throughout this work, since $\Delta t = 0$.

The composite likelihood of $\Delta t = 0$, $\Delta f$ events is calculated using two methods, depending on the experiment. **Method A** is used for the Haswell and Green Bank geographic spaced observations. **Method B** is used for the 164 beam transit test, using one dual orthogonal polarized telescope. Two methods are implemented, because the geographic spaced measurements use $\Delta f$ of associated pulse pairs to estimate AWGN likelihood, while the 164 beam transit test uses sorted decreasing SNR of $\Delta t = 0$, $|\Delta f| \leq 400$ Hz pulse pairs to estimate likelihood.

**Method A: Associated pulse pairs**

In this method, the protocol estimates the AWGN model likelihood of a set of energy burst pulses hypothesized to be associated with a $\Delta t = 0$, $|\Delta f| \leq 15.5$ Hz geographically spaced pulse pair event. The 15.5 Hz value is chosen based on metrology, residual Doppler compensation, and the observation of four $\Delta t = 0$, $\Delta f \approx 0$ Hz measurements, during two Haswell and Green Bank observation runs, August 15, 16, 2018 and April 2, 3, 2019, appearing anomalous within a single *RA* range, at likelihood quantified in text below **Tables 1 – 2**.

It may be hypothesized that the four $\Delta t = 0$, $|\Delta f| \leq 15.5$ Hz pulse pairs are the result of single-source transmitted signals received at both geographically spaced antennas, producing an SNR threshold crossing in each receiver. Associated pulse pairs are then sought.

A cumulative binomial probability of $\Delta t = 0$, $|\Delta f|$ associated spectral elements is calculated, using a single event probability equal to the $p_{0\Delta f\ AWGN}$ $(\Delta f)$ (14) of the most likely pulse pair among the candidate associated energy burst pulse pairs. The anomalous associated pulse pair candidates are selected as those having a $p_{0\Delta f AWGN}$ $(\Delta f)$ value less than 0.03.

A single $|\Delta f|$ is chosen to represent the $|\Delta f|$ values of $N_{\Delta f}$ associated pulse pairs, and has the maximum value of the $|\Delta f|$ values, i.e. most likely, measured within the set of $N_{\Delta f}$ associated pulse pairs.

The composite probability of the $\Delta t = 0$, $|\Delta f|$ associated pulse pairs is calculated as the sum of binomial density values, resulting in a binomial cumulative value of seeing $N_{\Delta f}$ or more associated pulse pairs, [17]

$$P_{0\Delta f AWGN} (\alpha_H, n, N_{\Delta f}) = \sum {}_n C_x \ \alpha_H{}^x (1 - \alpha_H)^{n-x} , \qquad (15)$$

where the cumulative summation in (15) has $x$ taking values $N_{\Delta f}$, $N_{\Delta f} + 1$, .. , $n$, where the 0 in the subscript in $P_{0\Delta f\ AWGN}(\alpha_H, n, N_{\Delta f})$ denotes $\Delta t = 0$, $\alpha_H$ is the trial event probability of the highest $|\Delta f|$ associated pulse pair, calculated using (14)

$$\alpha_H = \ln(2) \, / \, \Delta f \, / \, / \, \Delta f_{50\%} , \qquad (16)$$

$n$ is the number of trials, $N_{\Delta f}$ is the observed number of SNR threshold-crossing associated events at the same MJD, and $_n C_x$ is the binomial coefficient.

The number of trials $n$ in (15) is determined by selecting $\Delta t = 0$, $|\Delta f|$ pulse pairs, observed at the MJD of the $\Delta t = 0$, $\Delta f \approx 0$ Hz event, having an SNR greater than or equal to the lowest SNR among the associated $\Delta t = 0$, $|\Delta f|$ pulse pairs. **Method A** is illustrated graphically in **Figures 7, 12, 13**.

$\Delta f_{50\%}$ values in (16) were determined empirically, and analytically, from $|\Delta f|$ measurements of a receiver with an artificial noise source applied, and verified, within 3%, to the theoretically expected value of $\Delta f_{50\%}$ in AWGN, based on receiver instantaneous bandwidth and the average rate of SNR threshold crossings. Empirical and theoretical analysis establishes the $\Delta f_{50\%}$ value at 0.85 MHz at the receiver's

SNR threshold for telescope raw data file storage. The AWGN source model is assumed to be ergodic, so that $\Delta f_{50\%}$ applies within regions of the overall measured frequency band. The calculation of AWGN-model likelihood is expected to be minimally affected when relatively small portions of the spectrum are excised, due to machine RFI excision algorithms.

**Method B: Beam transits sorted SNR**

In this method, the experimental protocol calculates the binomial density of the count of $\Delta t = 0$, $|\Delta f| \leq 400$ Hz pulse pair events, measured within an *RA* window, among decreasing SNR-sorted $\Delta t = 0$, $|\Delta f| \leq 400$ Hz pulse pair events, having a higher SNR, in the same *RA* window.

**Method B** uses a calculated binomial probability density value, producing a likelihood function, *L*, a function of the number of trials, set by the SNR level of the $\Delta t = 0$, $|\Delta f| \leq 400$ Hz pulse pair event, and representing the binomial-based AWGN likelihood of seeing the observed count of $\Delta t = 0$, $|\Delta f| \leq 400$ Hz pulse pair events in the set of trials having higher values of SNR. The value of $\Delta f$ is measured and recorded, but not used in the likelihood function. The likelihood function values are compared across *RA* windows, assuming that *RA* windows are expected to have similar likelihood values, due to the AWGN hypothesis. The **Method B** measured likelihood is compared to the likelihood expected in AWGN, i.e. the measured likelihood of non-anomalous pulse pairs at the same number of trials. The significance of *RA* is examined in **APPENDIX D**.

## APPENDIX D
### *AWGN-model likelihood of RA anomalies*

The experiment in the current work entails telescope beam transit scans at -7.6° *DEC*. The AWGN model, as it describes SNR and $\Delta t$ $\Delta f$ measurements, detailed in the previous **APPENDIXES B and C**, does not naturally lend itself to explaining a theoretical correlation of pulse pair events to *RA*. Another method to calculate *RA* AWGN-model likelihood is required.

As described in **II. Hypothesis**, the AWGN model does not comprise astronomical natural sources, nor RFI. As a result of this hypothesized independence, the experimentally measured *RA* at the time of an SNR $\Delta t$ $\Delta f$ receiver event, is assumed to have a uniform probability density, within the overall observed *RA* range of the observation.

The probability $P_{RANGE}$ of an event occurring within a range of *RAs* under observation, may be calculated as

$$P_{RANGE} = (RA_{HIGH} - RA_{LOW}) / \ RA_{OBS} \qquad (17)$$

where $RA_{LOW}$ and $RA_{HIGH}$ set limits of a hypothesized *RA* range, and $RA_{OBS}$ is the overall range of *RA* that results from an observation.

Equation (17) calculates the probability of events in experiments where the observation of *RA* events within $RA_{OBS}$ is uniform.

In the 164 beam transit test, post processing four-hour duration telescope files, some *RA* values are sampled more often than others, producing a non-uniform probability across RA. The event probability within an *RA* range is estimated by assuming that almost all $\Delta t = 0$, $|\Delta f|$ elements are caused by AWGN, counting the number of $\Delta t = 0$, $|\Delta f| \leq 400$ Hz pulse pairs that fall within an *RA* range, over a wide range of SNRs, and dividing the count by the total number of $\Delta t = 0$, $|\Delta f| \leq 400$ Hz pulse pairs observed in the full observed *RA* range, in the same SNR wide range. Measured values of *RA* event probability are posted in V. **OBSERVATIONS *F. 164 beam transits of the Twenty-six Foot telescope.*** Comparison to the theoretical AWGN value is made.

### APPENDIX E
### *Bayesian inference*

Bayesian inference provides a framework to use a model

$$\theta \in [AWGN, RFI, ETI, ..]\qquad(18)$$

to calculate a posterior probability, where the $\theta$ value indexes a likelihood function, e.g. having index $\theta = AWGN$ in (18) and probability derived in (14),

$$p_0(\Delta f \mid \theta) = p_{0\Delta f\,AWGN}(\Delta f)\qquad(19)$$

where $p_0(\bullet|\bullet)$ is a conditional probability and subscript 0 indicates $\Delta t = 0$. $p_0(\theta)$ is the estimated probability that the model $\theta$ is valid in explaining prior $\Delta t = 0$ data, independent of new data, and is used in the Bayesian calculation of a posterior probability,

$$p_0(\theta \mid \Delta f) = p_0(\Delta f \mid \theta)\ p_0(\theta)\ /\ p_0(\Delta f)\qquad(20)$$

where $p_0(\Delta f)$ is a probability that new data is valid, normalizing the posterior probability.

$p_0(\theta \mid \Delta f)$ is the posterior probability that the model $\theta$ explains the observation of the newly acquired $\Delta f$ data, and may be used to subsequently provide a model's prior probability value, during a subsequent observation, providing new $\Delta f$ data [18].

The $p_0(\theta \mid \Delta f)$ posterior probability calculation is specific to each model, within each experimental hypothesis, and therefore cannot be used at one $\theta$ to infer the probability of other models explaining observational data. However, low values of $p_0(\theta \mid \Delta f)$ may be used to infer that one or more other models better explain the $\Delta f$ data, compared to an explanation based on the model $\theta$.

Bayesian data analysis is used in this work to update the hypothetical AWGN model validity in explaining hypothetical $\Delta t = 0$, $\Delta f$ events, using (20).